\documentclass[review,sort&compress]{elsarticle}

\usepackage[colorlinks,citecolor=blue,linktoc=all,linkcolor=cyan]{hyperref}
\usepackage{graphicx}

\usepackage[T1]{fontenc}
\usepackage{dsfont}               
\usepackage{mathrsfs}             
\usepackage{slashed}              
\usepackage{amsmath}
\usepackage{amssymb}
\usepackage{amsbsy}
\usepackage{amsfonts}

\usepackage{soul}

\usepackage{lipsum}
\usepackage{morefloats}
\usepackage{subfigure} 
\usepackage{rotating}
\usepackage{cancel}
\usepackage{mathtools}
\usepackage{color}
\usepackage{bbm}
\usepackage{dsfont}
\usepackage{bbold}
\usepackage{multirow}
\usepackage{ulem} 
\usepackage{wrapfig}
\usepackage{tikz-feynman}
\usepackage{environ}

\newcommand{\Mbare}{\stackrel{\circ}{M}_{\rm CQM}}

\newcommand{\olsi}[1]{\,\overline{\!{#1}}} 

\numberwithin{equation}{section}
\numberwithin{table}{section}
\numberwithin{figure}{section}

\journal{Progress in Particle and Nuclear Physics}

\usepackage{titlesec}
\usepackage{sectsty}
\titleformat{\section}{\normalfont\Large\bfseries}{\thesection}{1em}{}
\titleformat{\subsection}{\normalfont\large\bfseries}{\thesubsection}{1em}{}
\titleformat{\subsubsection}{\normalfont\normalsize\bfseries}{\thesubsubsection}{1em}{}

\begin{document}
	
	\begin{frontmatter}
		
		\title{Lowest-lying ${\frac{1}{2}}^-$ and ${\frac{3}{2}}^-$ $\Lambda_{Q}$ resonances: from the strange to the bottom sectors}

		
		\address[IFIC]{Instituto de F\'{\i}sica Corpuscular (centro mixto CSIC-UV), Valencia, Spain}
        \address[TUM]{Physik Department E62, Technische Universität München, Garching, Germany}
		\address[UESTC]{School of Physics, University of Electronic Science and Technology of China, Chengdu 611731, China}
		
		
		
		

            \author[IFIC]{J. Nieves}
		\ead{jmnieves@ific.uv.es}
		
		\author[IFIC,TUM]{A. Feijoo}
		\ead{edfeijoo@ific.uv.es}
		
		\author[IFIC]{M. Albaladejo}
		\ead{Miguel.Albaladejo@ific.uv.es}
		
		\author[UESTC]{Meng-Lin Du}
		\ead{du.ml@uestc.edu.cn}

		\begin{abstract}
We present a detailed study of the lowest-lying ${\frac{1}{2}}^-$ and ${\frac{3}{2}}^-$ $\Lambda_{Q}$ resonances both in the heavy quark (bottom and charm) and the strange sectors. We have paid special attention to the interplay between the constituent quark-model and chiral baryon-meson degrees of freedom, which are coupled using a unitarized scheme consistent with leading-order heavy quark symmetries. We show that the $\Lambda_b(5912)$ [$J^P=1/2^-$], $\Lambda_b(5920)$ [$J^P=3/2^-$] and the $\Lambda_c(2625)$ [$J^P=3/2^-$], and the $\Lambda(1520)$ [$J^P=3/2^-$] admitting larger breaking corrections, are heavy-quark spin-flavor siblings. They can be seen as dressed quark-model states with $\Sigma_{Q}^{(*)}\pi$ molecular components of the order of 30\%. The ${J^P=\frac{1}{2}}^-$ $\Lambda_c(2595)$ has, however, a higher molecular probability of at least $50$\%, and even values greater than 70\% can be easily accommodated. This is because it is located almost on top of the threshold of the $\Sigma_c\pi$ pair, which largely influences its properties. Although the light degrees of freedom in this resonance would be coupled to spin-parity $1^-$ as in the $\Lambda_b(5912)$, $\Lambda_b(5920)$ and $\Lambda_c(2625)$, the $\Lambda_c(2595)$ should not be considered as a heavy-quark spin-flavor partner of the former ones. We also show that the $\Lambda(1405)$ chiral two-pole pattern does not have analogs in the $\frac{1}{2}^-$ charmed and bottomed sectors, because the $ND^{(*)}$ and $N\olsi{B}{}^{(*)} $ channels do not play for heavy quarks the decisive role that the $N \olsi{K}$ does in the strange sector, and the notable influence of the bare quark-model states for the charm and bottom resonances. Finally, we predict the existence of two $\Lambda_b(6070)$ and two $\Lambda_c(2765)$ heavy-quark spin and flavor sibling odd parity states. 
		\end{abstract}
		
		
	\end{frontmatter}
	
	\newpage
	
	\thispagestyle{empty}
	\tableofcontents
	

	\newpage
	\section{Introduction and theoretical background}\label{sec:intro}
	\subsection%
           [Heavy quark symmetries, constituent quark models and the lowest-lying odd-parity $\Lambda^*_{b,c}$ resonances]%
 {\boldmath Heavy quark symmetries, constituent quark models and the lowest-lying odd-parity $\Lambda^*_{b,c}$ resonances}
	
 In the infinite quark mass limit ($m_Q\to \infty$), the spectrum of hadrons containing a heavy quark should show a $\text{SU}(2)$ pattern, because of the symmetry that Quantum Chromodynamics (QCD) acquires in that limit under arbitrary rotations of the spin of the heavy quark~\cite{Isgur:1991wq,Wise:1992hn,Neubert:1993mb}. This is known as heavy quark spin symmetry (HQSS).  In that case, the total angular momentum $j_q$ of the brown muck, which is the subsystem of the hadron apart from the heavy quark, is conserved, and hadrons with $J=j_q \pm \frac{1}{2}$ form a degenerate doublet. Additionally, in the $m_Q\to \infty$ limit, the dynamics of systems containing a single heavy quark becomes also independent of the flavor of the heavy quark. This is because the soft gluons which couple the heavy quark and the brown muck can only resolve distances much larger than $1/m_Q$. Hence, the light degrees of freedom ({\it ldof}\,) are blind to the flavor (mass) and spin orientation of the heavy quark. They only experience its color field, which extends over large distances because of confinement~\cite{Isgur:1991wq,Neubert:1993mb}.

\begin{table}[tbh]
    \centering
    \resizebox{1.\textwidth}{!}{
    \begin{tabular}{c|c|c|c||c|c|c}
    \hline\hline
         & \multicolumn{3}{c||}{$J^P= 1/2^-$} &\multicolumn{3}{c}{$J^P= 3/2^-$} \\\hline &&&&&&\\
         & $M-M^{1/2^+}_{\Lambda_Q}$ & $\Gamma$  & decay mode& $M-M^{1/2^+}_{\Lambda_Q}$ & $\Gamma$& decay mode \\&[MeV]&[MeV]&&[MeV]&[Mev]&\\ \hline
 $\Lambda_b(5912)$,  $\Lambda_b(5920)$&     $292.59\pm 0.24$ &$< 0.25$ (CL=90\%) & $\Lambda_b\pi\pi$ & $300.49\pm 0.24$ & $< 0.19$ (CL=90\%) &$\Lambda_b\pi\pi$\\  \hline            
 $\Lambda_c(2595)$,  $\Lambda_c(2625)$&     $305.79\pm 0.24$ & $2.6 \pm 0.6$& $\Lambda_c\pi\pi$ & $341.65\pm 0.13$ & $< 0.97$ (CL=90\%) &$\Lambda_c\pi\pi$\\  \hline
 $\Lambda(1405)$,  $\Lambda(1520)$ & $305^{+3}_{-2}{}^{+16}_{-20}$ & $38^{+16}_{-10}\pm 20$ & $\Sigma\pi$ & $403\pm 1$ & $16\pm 1$ & $N \olsi{K} $ (45\%), $\Sigma\pi$ (42\%) \\
             & $272\pm 9{}^{+11}_{-78}$& $230\pm 50{}^{+0}_{-140}$& $\Sigma\pi$ &$-$&$-$& $\Lambda\pi\pi (10\%)$ \\\hline\hline
    \end{tabular}
    }
    \caption{Properties of the lowest-lying $J^P= 1/2^-$ and $J^P= 3/2^-$ isoscalar resonances. Masses are referred to the ground state hyperon ($\Lambda_Q$)  with $J^P=1/2^+$ in each sector.  For the strange $\Lambda(1405)$ resonance, we assume a double pole structure, as discussed in the related note in the Review of Particle Physics (RPP)~\cite{ParticleDataGroup:2022pth}, and quote the results reported in Ref.~\cite{Guo:2012vv}, with the second set of errors accounting for the intervals covered by the different calculations (and their uncertainties) compiled in Table 83.1 of the RPP note. The rest of the data are taken from Ref.~\cite{ParticleDataGroup:2022pth}. }
    \label{tab:properties}
\end{table}

Constituent quark models (CQMs) predict a nearly degenerate pair of  P-wave $\Lambda_b^*$ excited states~\cite{Capstick:1986ter,Roberts:2007ni,Ebert:2007nw,Garcilazo:2007eh,Karliner:2008sv,Yoshida:2015tia}, with spin--parity $J^P=1/2^-$  and $3/2^-$, with masses similar to those of the isoscalar odd-parity $\Lambda_b(5912)$ and $\Lambda_b(5920)$ narrow resonances, observed by LHC in the $\Lambda_b^0\pi^+\pi^-$ mass spectrum~\cite{LHCb:2012kxf,LHCb:2020lzx,CMS:2020zzv}. The CDF II detector at the Fermilab Tevatron found also evidence for an excited resonance state~\cite{CDF:2013pvu}, with a mass of $5919.22\pm 0.76\,\text{MeV}$, consistent with the higher state $\Lambda_b(5920)$ reported by the LHCb and CMS experiments. Analogous predictions, as expected from heavy quark flavor symmetry (HQFS), are reported in the charm sector. These are then linked to the narrow $\Lambda_c(2595)$ and $\Lambda_c(2625)$ states observed 30 years ago in the $\Lambda_c^+\pi^+\pi^-$ mass spectrum by CLEO~\cite{CLEO:1994oxm}, E687~\cite{E687:1995srl} and ARGUS~\cite{ARGUS:1997snv}, and more recently by CDF~\cite{CDF:2011zbc}. Experimental properties of these isoscalar resonances are collected in Table~\ref{tab:properties}, where we can see that the mass splittings and partial decay widths of the bottom and charm states are, in good approximation,  independent of heavy quark flavor.
  
  The CQM study of $\Lambda_Q$ baryons carried out in Ref.~\cite{Yoshida:2015tia} accounts for excitations between the heavy quark and the brown muck as a whole ($\lambda-$mode),  and also excitations inside the brown muck ($\rho-$mode). When all quark masses are equal, $\lambda-$ and $\rho-$modes are degenerate~\cite{Yoshida:2015tia}. However, for singly-heavy baryons, the typical excitation energies of the $\lambda-$mode are smaller than those of the $\rho-$mode. This is because for singly charm or bottom baryons, the interactions between the heavy quark and the brown muck are more suppressed than between the light quarks~\cite{Isgur:1991wr, Yoshida:2015tia}. Thus, one should expect the $\lambda$ excitation modes to become dominant for low-lying states of singly heavy-quark baryons. Within this picture, $\Lambda^{\rm CQM}_b(5912)$ and $\Lambda^{\rm CQM}_b(5920)$ and $\Lambda^{\rm CQM}_c(2595)$ and $\Lambda^{\rm CQM}_c(2625)$ resonances correspond to the members of the HQSS--doublet associated to   $(\ell_\lambda=1,\ell_\rho=0)$, with total spin $S_q=0$ for the {\it ldof}, leading to a  spin-flavor-spatial symmetric wave-function for the light isoscalar diquark subsystem inside of the $\Lambda_{b,c}^*$ baryon. The total spins of these states are the result of coupling the orbital-angular momentum $\ell_\lambda$ of the brown muck --with respect to the heavy quark-- with the spin ($S_Q$) of the latter. Thus both $\Lambda^{\rm CQM}_b(5912)$ and $\Lambda^{\rm CQM}_b(5920)$ or $\Lambda^{\rm CQM}_c(2595)$ and $\Lambda^{\rm CQM}_c(2625)$ states are connected by  a simple rotation of the heavy-quark spin, and these resonances will be degenerate in the heavy-quark limit.\footnote{The lowest-lying $\rho-$mode $(\ell_\lambda=0,\ell_\rho=1)$ gives rise to two $\frac12^-$ and also two $\frac32^-$ multiplets of $\Lambda^*_{b,c}$ states, together with an additional $\frac52^-$ $\Lambda^*_{b,c}-$excited state, significantly higher in the spectrum~\cite{Yoshida:2015tia}. Note that the isoscalar light diquark could have $0^-$, $1^-$, and $2^-$ quantum-numbers, resulting from the coupling of the spin, $S_q=1$, and the orbital-angular momentum, $\ell_\rho=1$, of the light quarks. In the heavy quark limit all the baryons with the same light diquark $j_q^\pi$ configuration will be degenerate~\cite{Isgur:1991wr}.} For finite $m_Q\gg \Lambda_{\rm QCD}$, these symmetries become approximate ones which are broken by effects of order ${\cal O}(\Lambda_{\rm QCD}/m_Q)$  and ${\cal O}(\alpha_s(m_{Q})/\pi)$. 

On the other hand, the $\Lambda^{\rm CQM}_{c,b}(2595,5912) \to\Sigma_{c,b}\pi  \to  \Lambda_{c,b}\pi\pi$ and $\Lambda^{\rm CQM}_{c,b}(2625,5920) \to \Sigma^*_{c,b}\pi  \to  \Lambda_{c,b} \pi\pi $ decays respect HQSS, and hence one might expect sizable widths for these resonances, unless these transitions are kinematically suppressed. This latter scenario seems plausible, given the involved masses and thresholds, which are shown below in Fig.~\ref{fig:espectro}. Indeed, the works of Refs.~\cite{Nagahiro:2016nsx, Arifi:2017sac} for the charm sector find widths for the CQM $(\ell_\lambda=1,\ell_\rho=0)$ states ($j_q^\pi=1^-$)  predicted in Ref.~\cite{Yoshida:2015tia} consistent with data.

The strange sector CQMs, though predict reasonably well the $J^P=3/2^-$ $\Lambda(1520)$ state, cannot naturally describe the $\Lambda(1405)$ resonance, which is typically predicted above 1.5 GeV~\cite{Yoshida:2015tia}.  This is commonly accepted as a hint of the importance of the effects produced by S-wave Goldstone boson-baryon loops~\cite{Kaiser:1995eg, Oset:1997it,Oller:2000fj,Garcia-Recio:2002yxy}, driven by the approximate chiral symmetry (CS) of QCD.

\subsection{Unitary chiral perturbation theory and hadroni-molecules}
\label{sec:uchpt}
CS and its pattern of spontaneous and explicit breaking allow one to construct an effective field theory, which has proven to be successful in describing low-energy phenomena involving hadrons made of $u, d$, and $s$ quarks. CS is deduced from QCD in the opposite limit to heavy-quark symmetries, \textit{i.e.}, when the quark masses are much smaller than the confinement scale $\Lambda_{\rm QCD}$~\cite{Gasser:1983yg, Gasser:1984gg, Pich:1995bw}. However, by construction, chiral perturbative theory (ChPT)  cannot describe hadron resonances given their non-perturbative character, for which unitarity plays a crucial role. Moreover, resonances are in general located at energies above the range of applicability of the perturbative expansion. It has been shown that by unitarizing the perturbative coupled-channel amplitudes one can greatly extend the region of application of ChPT.\footnote{Several frameworks have been proposed to unitarize the amplitudes, though the most common and successful are the inverse amplitude method and the solution of the Bethe-Salpeter equation. In this latter case, several renormalization procedures have been also employed, differing mostly in the treatment of the off-shell effects and the finite part of the leading ultraviolet  terms. In general, the different methods give similar results for the lowest-lying resonances.} This approach, commonly referred as Unitary Chiral Perturbation Theory (UChPT), has received much attention and provided a plethora of interesting results~\cite{Truong:1988zp,Dobado:1989qm,Dobado:1992ha,Kaiser:1995eg,Dobado:1996ps,Hannah:1997ux,Oset:1997it, Oller:1997ng, Oller:1997ti,Nieves:1998hp,Oller:1998hw,Oller:1998zr,Nieves:1999bx, Nieves:2000km,Oller:2000fj,GomezNicola:2000wk,Nieves:2001de,GomezNicola:2001as,Lutz:2001yb,Nieves:2001wt,Hyodo:2002pk,Garcia-Recio:2002yxy, Kolomeitsev:2003kt,Lutz:2003fm,Jido:2003cb,Garcia-Recio:2003ejq, Sarkar:2004jh, Borasoy:2005ie, Roca:2005nm, GomezNicola:2007qj,Toki:2007ab,Hyodo:2007jq,Hanhart:2008mx,Hyodo:2008xr,Albaladejo:2008qa,Nebreda:2010wv,Garcia-Recio:2010enl,Bruns:2010sv,Nieves:2011gb,Hyodo:2011ur,Guo:2012vv,Ikeda:2012au,Mai:2014xna,Ledwig:2014cla,Pelaez:2015qba,Kamiya:2016jqc, Lu:2022hwm}. In particular, many meson-meson and meson-baryon resonances and bound states can be naturally described within UChPT. These states are then interpreted as hadron-molecules having a ``dynamical nature''. In other words, they are not genuine $q\bar q$ or $qqq$ states, but are mainly built out of their meson-meson or meson-baryon components. One possible way to distinguish these two pictures is to study the dependence on the number of colors of the resonance masses and widths ~\cite{Pelaez:2003dy,Pelaez:2006nj,Garcia-Recio:2006uwg,Hyodo:2007np,Roca:2008kr,Geng:2008ag, Nieves:2009ez, Nieves:2009kh,Nieves:2011gb,Pelaez:2015qba}.

UChPT studies of the $\Lambda(1405)$ region (see for instance the first works on the subject of Refs.~\cite{Oset:1997it, Oller:2000fj,Garcia-Recio:2002yxy,Jido:2003cb,Garcia-Recio:2003ejq} and the dedicated note in the RPP~\cite{ParticleDataGroup:2022pth})  find two poles of the scattering amplitude in the complex energy plane between the $N \olsi{K}$ and $\Sigma\pi$ thresholds.  The two poles are in the  $\Sigma\pi$ unphysical Riemann sheet, and some details are given in Table~\ref{tab:properties}. The spectrum in experiments exhibits one effective resonance shape, while the existence of two poles results in the reaction-dependent lineshape.\footnote{The GlueX collaboration has recently reported the measurement of the $\gamma p \to K^+ \pi^0 \Sigma^0$ photo-production process \cite{Wickramaarachchi:2022mhi}, showing hints of a two-pole structure in the $\pi^0 \Sigma^0$ invariant mass spectrum for the first time.} The origin of this two-pole structure is attributed to the two attractive channels of the leading order Weinberg-Tomozawa (WT) chiral interaction in the $\text{SU}(3)$ basis (singlet and octet)  or in the isospin basis ($N \olsi{K}$ and $\Sigma\pi$). Analysis of the lattice QCD (LQCD) data also supports this picture~\cite{Menadue:2011pd,Engel:2013ig,Hall:2014uca,Liu:2016wxq,Gubler:2016viv,Molina:2015uqp,Pavao:2020zle}.

The total angular momentum and parity ($j_q^\pi$) of the {\it ldof} of the  $\Sigma_b\pi$ ($\Sigma_c\pi$) and $\Sigma^*_b\pi$ ($\Sigma_c^*\pi$) pairs in S-wave are $1^-$, as in the $\Lambda_b(5912)$ and $\Lambda_b(5920)$ ($\Lambda_c(2595)$ and $\Lambda_c(2625)$) resonances.  It is therefore appropriate to investigate the role played by the $\Sigma_Q^{(*)}\pi$ baryon-meson pairs in the generation of these odd-parity bottomed and charmed resonances, especially in the $\Lambda_c(2595)$ case, since it is located very close to the $\Sigma_c\pi$ threshold (1 MeV below or four MeV above depending on the charges in the channel). Moreover, we have seen above the importance of the $\Sigma\pi$ channel in the dynamics of the two states associated to the PDG $\Lambda(1405)$  resonance. These observations lead naturally to  possible molecular descriptions of these lowest-lying odd parity charmed/bottomed baryon states, which should show up as poles in the $\Sigma_Q^{(*)}\pi$ $T-$matrices, fulfilling exact unitary. 

The interactions between the ground-state $\frac{1}{2}^+$ and $\frac{3}{2}^+$ singly charmed and bottomed baryons and the pseudo-Nambu-Goldstone bosons are constrained by CS and a perturbation theory for hadrons containing a heavy quark, \textit{i.e.} heavy quark ChPT (HHChPT), can be consistently constructed~\cite{Wise:1992hn,Casalbuoni:1996pg}. Similarly, as we discussed above, the unitarization of the HHChPT coupled-channel amplitudes greatly extends the region of application of the perturbative scheme and allows for the description of non-perturbative phenomena, like resonances and bound states. A detailed treatment, within this framework and an additional discussion on the effects of the next-to-leading-order chiral potentials, of the $\Lambda_b(5912)$ and $\Lambda_b(5920)$, and  $\Lambda_c(2595)$ and $\Lambda_c(2625)$ resonances was carried out in Ref.~\cite{Lu:2014ina}. Dynamically generated states are naturally found within this unitarized chiral approach but with masses generally higher than those of the physical resonances. Actually, large ultraviolet (UV) cutoffs of $2.17$, $2.60$, $1.35$, and $2.13\,\text{GeV}$ are needed to move the molecular states down to the positions of the $\Lambda_b(5912)$, $\Lambda_b(5920)$,  $\Lambda_c(2595)$ and $\Lambda_c(2625)$, respectively. These large UV scales point out,  considering the arguments given in Refs.~\cite{Guo:2016nhb, Albaladejo:2016eps}, to the existence of some relevant degrees of freedom  (for instance the CQM compact states mentioned above)  that are not properly accounted for in Ref.~\cite{Lu:2014ina}. The $\Lambda_c(2595)$ requires the smallest UV cutoff, though it is still far beyond the edge of acceptable values ($\sim 0.5$--$1.0\,\text{GeV}$), hinting to the probable importance of the $\Sigma_c\pi$ component in its dynamics.

\subsection{Extended molecular schemes}
It is appealing to investigate the $N D$ and $N \olsi{B}$ channels within the hadronic picture of these odd-parity heavy-flavor resonances, given the role played by the analog $N \olsi{K}$ channel in the case of the $\Lambda(1405)$.  The pioneering work of Ref.~\cite{Tolos:2004yg} for the $\Lambda_c(2595)$ resonance exploited the possible analogy between the charmed and strange sectors. The interactions were obtained from chirally motivated Lagrangians upon replacing the $s$ quark with the $c$ quark. Though in this way the $DN$ channel was accounted for, the HQSS counterpart $N D^*$  was not considered. The $D$ and $D^*$ (or $\olsi{B}$ and $\olsi{B}{}^*$) are HQSS partners and both mesons must be treated equally in the heavy quark limit, since they are degenerate and are connected by a spin-rotation of the heavy quark that leaves unaltered the QCD Hamiltonian in that limit. It is natural then to also consider the $N D^*$ and $N \olsi{B}{}^* $ channels in the molecular scheme, which allow one to construct S-wave $N D^*$ and $ N \olsi{B}{}^* $ pairs coupled not only to spin-parity $\frac{1}{2}^-$, but also to $\frac{3}{2}^-$. 

Subsequent works~\cite{Hofmann:2005sw, Mizutani:2006vq, Hofmann:2006qx,Jimenez-Tejero:2009cyn} introduced some
improvements, but they failed to provide a scheme fully consistent with HQSS. In all cases, the $\Lambda_c(2595)$,  or the $\Lambda_c(2625)$ when studied, could be dynamically generated after a convenient tuning of the low energy constants (LECs) needed to renormalize the UV divergences resulting from the baryon-meson loops.  

The first molecular description of the $\Lambda_c(2595)$ and $\Lambda_c(2625)$ resonances, using interactions fully consistent with HQSS, was proposed in Refs.~\cite{Garcia-Recio:2008rjt, Romanets:2012hm}. In these works, a consistent ${\rm SU(6)}_{\rm lsf} \times {\rm SU(2)}_{\rm HQSS}$ extension of the chiral WT $N \pi $ Lagrangian (where ``lsf'' stands for light-spin-flavor symmetry) was derived,\footnote{The extension of the WT interaction to include light-vector mesons~\cite{Garcia-Recio:2005elc,Gamermann:2011mq} is just a model, which does not rely on CS, since the latter symmetry does not fix the interaction between vector mesons and baryons. On the other hand, HQSS connects vector and pseudoscalar mesons which contain a charm (heavy) quark. It does not tell anything about mesons made out of light quarks. Nevertheless, it is clearly appealing to have a model for four flavors and for pseudoscalar and vector mesons which reduces to the WT interaction in the sector where Goldstone bosons are involved and which incorporates HQSS when charm (heavy) quark hadrons participate.} and was used as a kernel of the Bethe-Salpeter equation (BSE). The model is supplemented by a pattern of symmetry-breaking corrections and uses the particular renormalization scheme (RS) proposed in Refs.~\cite{Hofmann:2005sw,Hofmann:2006qx} (see also for instance Sect. II.B of Ref.~\cite{Nieves:2019nol}). In what follows, we will refer to this RS as  SC$\mu$. In Refs.~\cite{Garcia-Recio:2008rjt, Romanets:2012hm} two states with $J^P =\frac{1}{2}^-$ were dynamically generated in the region of $2595\,\text{MeV}$. The first one, identified with the $\Lambda_c(2595)$ resonance, was narrow and it strongly coupled to $DN$ and especially to $N D^*$, with a small coupling to the open $\Sigma_c\pi$ channel. Its light-quark wave-function had a large $j_q^\pi=0^-$ component that, when coupled to the spin ($S_Q=\frac{1}{2}$) of the charm quark, gives a total $J^P = \frac{1}{2}^-$ for the $\Lambda_c(2595)$. Since the transition of the dominant $j_q^\pi=0^-$ term of the $\Lambda_c(2595)$ to the final $\Sigma_c\pi$ state is forbidden by HQSS, this mechanism will act in addition to any possible kinematical suppression. The second $J^P =\frac{1}{2}^-$ state found in Refs.~\cite{Garcia-Recio:2008rjt, Romanets:2012hm} was quite broad due to its sizable coupling to the $\Sigma_c\pi$ channel, and was analogous, in the charm-sector, to the second resonance in the chiral two-pole structure of the $\Lambda(1405)$~\cite{Oller:2000fj,Garcia-Recio:2002yxy,Hyodo:2002pk,Jido:2003cb,Garcia-Recio:2003ejq,Hyodo:2011ur, Kamiya:2016jqc}. On the other hand, a $J^P=\frac{3}{2}^-$ state was generated mainly by the $N D^*$--$\Sigma_c^*\pi$ coupled-channel dynamics. It would be the charm counterpart of the $\Lambda(1520)$, and it was argued that it could be identified as the $\Lambda_c(2625)$ resonance. In addition, the second, broad, $J^P =\frac{1}{2}^-$ resonance in the region of the $\Lambda_c(2595)$, with a large $j_q^\pi=1^-$ {\it ldof} component, could be naturally identified with the HQSS partner of the $\Lambda_c(2625)$ $(J^P=\frac{3}{2}^-)$, since both states will have the same brown muck configuration in the heavy-quark limit. The same ${\rm SU(6)}_{\rm lsf} \times {\rm SU(2)}_{\rm HQSS}$ model also dynamically generates the $\Lambda_b(5912)$ and $\Lambda_b(5920)$ narrow resonances, which turn out to be HQSS partners, thereby explaining naturally their approximate mass degeneracy~\cite{Garcia-Recio:2012lts}, and was employed to study other sectors involving heavy quarks as well.\footnote{The extension of the model to the hidden charm  sector was carried out in Ref.~\cite{Garcia-Recio:2013gaa}, and more recently it was shown~\cite{Nieves:2017jjx} that some (probably at least three) of the narrow $\Omega_c^*$ states recently observed by LHCb~\cite{LHCb:2017uwr} in the $\Xi_c^+K^-$ spectrum in $pp$ collisions can be also dynamically generated within the same scheme. Finally, $\Xi_c$ and $\Xi_b$ excited states in the ${\rm SU(6)}_{\rm lsf}\times\text{HQSS}$ model were also discussed in Ref.~\cite{Nieves:2019jhp}, paying special attention to the excited $\Xi_b(6227)$ state reported by LHCb in the $\Lambda_b^0 K^-$ and $\Xi_b^0 \pi^-$ invariant mass spectra also in $pp$ collisions \cite{LHCb:2018vuc}, and to the $\Xi_c(2930)$ state observed in its decay to $\Lambda_c^+ K^-$ in the $B^- \rightarrow K^- \Lambda_c^+ \overline{\Lambda}_c^-$ reaction  measured by Belle~\cite{Belle:2017jrt}. In the latter two works about the $\Omega_c^*$, $\Xi_c$, and $\Xi_b$ excited states, results adopting a UV cutoff were obtained, in addition to those found with the SC$\mu$ RS.}

Several $\Lambda_c^*$ poles were also obtained in the approach followed in
Ref.~\cite{Liang:2014kra}. There, the interaction of $N D$ and $N D^*$ states, together with their coupled
channels are considered by using an extension of the SU(3) local hidden gauge (ELHG) formalism from the light meson sector~\cite{Bando:1984ej,Bando:1987br,Meissner:1987ge} to four flavors. The scheme also respects leading order (LO) HQSS constraints~\cite{Xiao:2013yca} and the BSE amplitudes are renormalized by fine-tuning several UV sharp-cutoffs to the pole positions. As in Refs.~\cite{Garcia-Recio:2008rjt,Romanets:2012hm}, a two-pole structure for the $\Lambda_c(2595)$ was also found in Ref.~\cite{Liang:2014kra}, with the $N D^*$ channel playing a crucial role in its dynamics. This is a notable difference from the situation in the strange sector, where the analog $N \olsi{K}{}^* $ channel is not even considered in most of the studies of the $\Lambda(1405)$, because of the large $\olsi{K}{}^*$--$\olsi{K}$ mass splitting. In addition, the $\Lambda_c(2625)$ also turns out to be essentially a $N D^*$ state, while the isoscalar $\Sigma_c^*\pi$ is treated as a single channel, which gives rise to a further broad state ($\Gamma\sim 100\,\text{MeV}$) in the region of $2675\,\text{MeV}$, which is not related to the $\Lambda_c(2625)$ in Ref.~\cite{Liang:2014kra}. Some of the newly observed $\Omega_c$, $\Xi_c$, and $\Xi_b$ excited states are also described as hadronic molecules within the ELHG scheme~\cite{Debastiani:2017ewu,Liang:2017ejq, Debastiani:2018adr,Yu:2018yxl}. The main differences between the ${\rm SU(6)}_{\rm lsf} \times {\rm SU(2)}_{\rm HQSS}$ and ELHG schemes are discussed with some details in  Sec. 2.2 of Ref.~\cite{Nieves:2019kdh}. In addition to some differences between the interactions not restricted by CS, these are related to the renormalization of the UV divergences and the fine-tuning of their finite parts.

\subsection{Interplay between baryon-meson and bare CQM degrees of freedom}
Neither the ${\rm SU(6)}_{\rm lsf} \times {\rm SU(2)}_{\rm HQSS}$ model, nor the ELHG scheme considers 
the interplay between $\Sigma_c^{(*)}\pi$-$ND^{(*)}$ baryon-meson degrees of freedom and bare P-wave $c ud$ quark-model states. This is as unjustified as it was to neglect the baryon-meson effects in the CQM approaches.

CQM degrees of freedom can be taken into account in hadron scattering schemes by considering an additional energy-dependent interaction~\cite{Baru:2010ww, Cincioglu:2016fkm}, driven by a pole in the baryon-meson tree-level amplitudes located at the bare mass ($\Mbare$) of the CQM state. At energies far enough from $\Mbare$, the contribution of the CQM degrees of freedom can be possibly accounted for an appropriate LEC (induced by the UV regulator of the loops) in the unitarized baryon-meson amplitude. However, such contribution becomes more important for energies approaching $\Mbare$, and its energy dependence might then not be safely ignored. 

\begin{figure}[t]
\begin{center}
\includegraphics[width=1\textwidth]{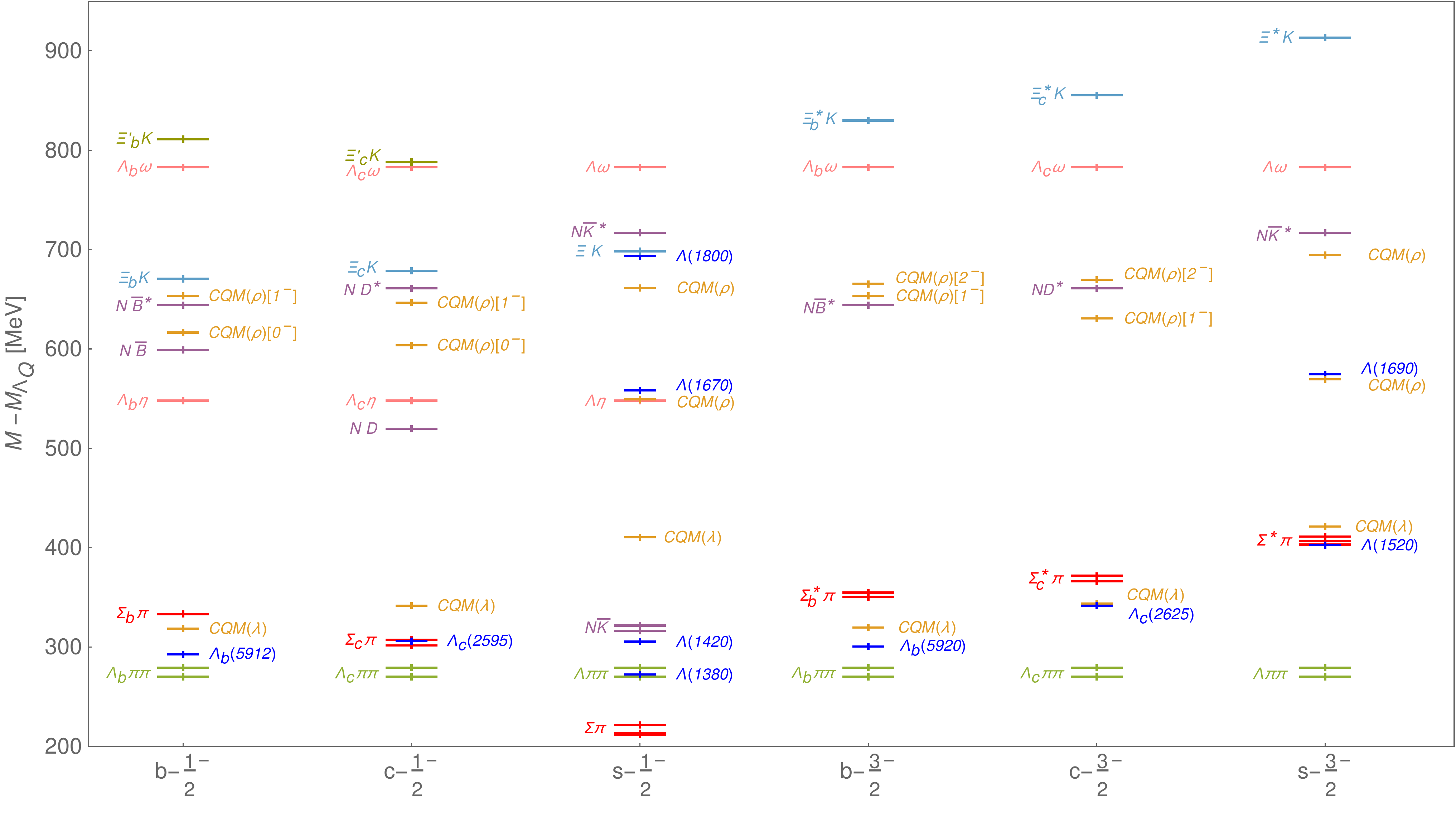}
\end{center}
\caption{Physical resonance masses,  predictions of the $\lambda$- and $\rho$-excitations obtained in the CQM of Ref.~\cite{Yoshida:2015tia} and positions of different thresholds, in the bottom ($b$), charm ($c$) and strange ($s$) sectors and spin-parities $J^P=1/2^-$ and $3/2^-$.  All energies are in MeV and are referred to the mass of the corresponding ground state hyperon ($\Lambda_b$, $\Lambda_c$ or $\Lambda$)  with $J^P=1/2^+$. The total angular momentum and parity of the {\it ldof} are given for the CQM  bottom and charm $\rho-$mode states, for which HQSS-breaking is expected to be negligible, while $j_q^\pi$ is $1^-$ for $\lambda-$excitations.  }\label{fig:espectro}
\end{figure}

Paying attention to the different energy-splittings displayed in Fig.~\ref{fig:espectro}, we observe a clear variation-pattern between the relative positions of the resonances and different thresholds, when moving from the bottom to the charm and finally to the strange sectors. Moreover from the figure, one would expect that for the dynamics of the lowest-lying resonances odd-parity $\Lambda_b(5912)$, $\Lambda_b(5920)$, $\Lambda_c(2595)$, $\Lambda_c(2625)$, $\Lambda(1405)$ and  $\Lambda(1520)$:
\begin{itemize}
    \item The lowest $\lambda-$mode CQM state should play a significant role in all cases,  except probably for the $\Lambda(1405)$.
    \item In the charm and bottom sectors, among the chiral channels considered in Ref.~\cite{Lu:2014ina}, only the  $\Sigma_{Q}^{(*)}\pi$ ones should be important.
    \item Although the $N \olsi{K} $ channel must be considered to understand/generate the chiral two-pole pattern of the $\Lambda(1405)$, for the other resonances, the explicit inclusion of nucleon channels ($N \olsi{B}{}^{(*)}, N D^{(*)}$ and $N \olsi{K}{}^{*} $) should not be necessary, since they are at least $200$--$300\,\text{MeV}$ above the nominal position of the states. Appropriate LECs should  be able to take into account, with sufficient precision, the contributions of these high-energy channels, in the spirit of an effective approach only reliable in a limited energy range.
\end{itemize}

In relation to the last point above, we should mention that  in the ${\rm SU(6)}_{\rm lsf} \times {\rm SU(2)}_{\rm HQSS}$ model, the $N D{}^{(*)}$ or $N \olsi{B}{}^{(*)}$ channels acquire relevance in the $J^P=\frac{1}{2}^-$ sector only because of the SC$\mu$ RS. Such RS introduces UV cuts, sometimes very different and also far from the physical-motivated values of around $650$--$750\,\text{MeV}$, depending on the channel (see Sec. 2.B of Ref.~\cite{Nieves:2019nol}). Actually, in the ${\rm SU(6)}_{\rm lsf} \times {\rm SU(2)}_{\rm HQSS}$ model, when a common  UV cutoff is used for all channels, the two-pole pattern for the $\Lambda_c(2595)$ ($\Lambda_b(5912)$) resonance disappears, and one would essentially recover the spectrum reported in Ref.~\cite{Lu:2014ina}, with additional states in the vicinity of the $N D^{(*)}$ ($N \olsi{B}{}^{(*)}$) thresholds~\cite{Nieves:2019nol}. In the $J^P=\frac{3}{2}^-$ sector, neither the SC$\mu$, nor the common UV RS provides a good description of the $\Lambda_c(2625)$ and $\Lambda_b(5920)$ resonances. Both schemes give rise to broad states around the $\Sigma_{b,c}^*\pi$ threshold, its identification with the physical resonances being very doubtful. Analogously, in the ELHG scheme the hardly understandable importance of the the $N D$ and $N D^* $  channels in the dynamics of the $\Lambda_c(2595)$ and $\Lambda_c(2625)$ resonances is also likely an artifact produced by a cumbersome RS, with several UV cutoffs adjusted to reproduce poles that could be associated to these states, in a context where the possible influence of CQM degrees of freedom is ignored. 

In this work, we study the strange, charm, and bottom lowest-lying ${\frac{1}{2}}^-$ and ${\frac{3}{2}}^-$ $\Lambda_{(Q)}$ resonances using a model which considers the interplay between the nearest baryon-meson and bare CQM degrees of freedom. For the former ones, as discussed above, though the $N \olsi{K}$ channel must be considered to understand the $\Lambda(1405)$ chiral two-pole pattern, for the other resonances, we do not include the $N \olsi{K}{}^*, N D^{(*)}$ and $N \olsi{B}{}^{(*)}$ nucleon-channels. Therefore, we could determine the baryon-meson interactions only from the scattering of pion and antikaon Goldstone-bosons off baryons. As in Ref.~\cite{Nieves:2019nol}, we employ a common UV cutoff in all channels to renormalize the hadron-loops. Such procedure is consistent with HQSS~\cite{Lu:2014ina} and prevents from abnormally large influence of high energy channels. In addition in the heavy-quark sectors, we constrain the couplings between CQM and meson-baryon states using HQSS. Some of the results obtained here for the $\Lambda_c(2595)$ and $\Lambda_c(2625)$ were already obtained in some of the cases discussed in Ref.~\cite{Nieves:2019nol}, which was the first approach to combine CQM and baryon-meson degrees of freedom to study these two resonances.  

 We will show that the $\Lambda(1405)$ chiral two-pole pattern does not have analog in the $\frac{1}{2}^-$ charmed and bottomed sectors, because the $ND^{(*)}$ and $N \olsi{B}{}^{(*)}$ channels do not play for the heavy quarks the decisive role played by the $N \olsi{K}$ in the strange sector, and furthermore, the large influence of the bare CQM states for the charm and bottom resonances is not as important in the strange ${\frac{1}{2}}^-$ sector. Moreover, we will also discuss the great importance of taking into account the chiral $ \Sigma_{c,b}^{(*)}\pi$  channels and their interplay with the CQM degrees of freedom.

The double-pole structure for the $\Lambda_c(2595)$, with a quite broad second pole, cannot be firmly discarded by the spectra reported by the CLEO, E687, ARGUS and CDF experiments~\cite{CLEO:1994oxm,E687:1995srl,ARGUS:1997snv,CDF:2011zbc}.   The  $\Lambda_b \to \Lambda_c(2595) \ell^-\bar\nu_\ell$ and $\Lambda_b \to \Lambda_c(2625) \ell^-\bar\nu_\ell$ reactions, assuming a single pole structure for the  $\Lambda_c(2595)$ and a  $j^\pi_q=1^-$ configuration for both this resonance and the  $\Lambda_c(2625)$, have been studied using heavy quark effective theory~\cite{Roberts:1992xm, Leibovich:1997az,Baccouche:2001snw,Nieves:2019kdh}, non-relativistic and semi-relativistic CQMs~\cite{Pervin:2005ve}, and in Refs.~\cite{Meinel:2021rbm,Papucci:2021pmj,Du:2022ipt} using the LQCD form-factors obtained in Refs.~\cite{Meinel:2021rbm, Meinel:2021mdj}. A double-pole for the $\Lambda_c(2595)$ would lead to significantly different results for the total and partial integrate rates of the exclusive semileptonic $\Lambda_b$ decay into $\Lambda_c(2595)$~\cite{Liang:2016exm,Nieves:2019kdh}. 

Recently, it has also been suggested that  measurements of the semileptonic $\Lambda_b \to \Lambda_c(2595), \Lambda_c(2625)$ decays by LHCb could be used to perform precise lepton flavor universality  tests~\cite{Boer:2018vpx, Gutsche:2018nks,Nieves:2019kdh,Papucci:2021pmj,Du:2022ipt},  comparing branching fractions with $\tau-$ or $\mu-$leptons in the final state. The analyses in the latter works assumed that both excited charmed baryons form a doublet under HQSS, and therefore it neither contemplated the possibility that the narrow $\Lambda_c(2595)$ might not be the HQSS partner of the $\Lambda_c(2625)$, nor that it could contain a non-negligible $j_q^\pi=0^-$ component, as it occurs in some of the molecular descriptions of this resonance reviewed above.\footnote{We should mention that the recent study carried out in Ref.~\cite{Du:2022rbf}, using the semileptonic LQCD form-factors obtained in Refs.~\cite{Meinel:2021rbm, Meinel:2021mdj}, gives certain support, or at least does not contradict, that the  $\Lambda_c(2595)$ and $\Lambda_c(2625)$ resonances might form the lowest-lying  HQSS $j_q^\pi=1^-$ doublet.} It is therefore timely and of the utmost interest to test the HQSS doublet assumption for the $\Lambda_c(2595)$ and $\Lambda_c(2625)$.  In this context, the combined analysis of the strange, charm, and bottom lowest-lying ${\frac{1}{2}}^-$ and ${\frac{3}{2}}^-$ $\Lambda_{Q}$ resonances provides robust theoretical hints on the actual nature of the $\Lambda_c(2595)$ and $\Lambda_c(2625)$ states.

The paper is organized as follows. After this extensive Introduction, where we have also reviewed the different interpretations available in the literature of the $\Lambda_Q$ states, in Sec.~\ref{sec:tmatrix} we describe the approach used in this work to obtain the unitarized  $T-$matrix accounting for both baryon-meson and CQM degrees of freedom. Next, we relate the different resonance observables with the properties of the poles of the $T-$matrix in the complex plane. In Sec.~\ref{sec:spectrum}, we describe in detail the spectrum obtained in this work. Subsections~\ref{sec:dq-bottom} and \ref{sec:bottom-poles} are devoted to the bottom sector. The properties of the $\Lambda_b(5912)$ and $\Lambda_b(5920)$ are discussed in the former subsection and they are used to constrain the UV cutoff and the LEC that controls the interplay between CQM and chiral degrees of freedom. Higher resonances, related to the $\Lambda_b(6070)$, are presented in Subsec.~\ref{sec:bottom-poles}. The $\Lambda_c(2595)$, $\Lambda_c(2625)$ and $\Lambda_c(2765)$ charmed resonances are discussed in Sec.~\ref{sec:charm}, while the strange sector is studied in Sec.~\ref{sec:strange}. Finally, the main conclusions of this work are collected in Sec.~\ref{sec:concl}.   
%
%
	
\section{\boldmath $T-$matrix model, poles and residues}
\label{sec:tmatrix}
To construct the interactions, evaluate the loop functions and look for poles in the first and second Riemann sheets (FRS and SRS) on the unitarized amplitudes, we use charged averaged hadron masses,\footnote{Except for the pion mass, for which we take that of the charged $\pi^{\pm}$ pions. This is to guarantee that in the bottom sector, where the masses of the neutral $\Sigma_b^{0}$ and $\Sigma_b^{*0}$ baryons are not given in the RPP, the known  $\Sigma_b^{(*)\pm}\pi^\mp$ thresholds are correctly evaluated. } which are compiled in Table~\ref{tab:numval}. This approximation induces systematic errors on the position of the dynamically generated states of the order of a few MeV in some cases. For instance, $M_{\Sigma_b^-}-M_{\Sigma_b^+} \sim 5$ MeV, which is an isospin splitting similar to that found between charged and neutral pion masses. 
\begin{table}[tbh]\label{tab:numval}
    \centering
    \resizebox{1.\textwidth}{!}{
    \begin{tabular}{c|c|c|c||c|c|c}
    \hline\hline
       & \multicolumn{3}{c||}{$J^P= 1/2^-$} & \multicolumn{3}{c}{$J^P= 3/2^-$} \\\hline 
         & $\Mbare$ \cite{Yoshida:2015tia}&  $M_{\Sigma_Q}$  & $(M,\Gamma)$ RPP~\cite{ParticleDataGroup:2022pth} & $\Mbare$ \cite{Yoshida:2015tia} &  $M_{\Sigma_Q^*}$ & $(M,\Gamma)$ RPP~\cite{ParticleDataGroup:2022pth}\\&[MeV]&[MeV]&[MeV]&[MeV]&[MeV]&[MeV]\\ \hline
 bottom & 5938     &$5813.1\pm 2.5$ & $(5912.19\pm 0.17, <0.25)$ & 5939 & $5832.5\pm 2.2$ & $(5920.09\pm 0.17,< 0.19)$ \\  \hline            
 charm & 2628     & $2453.5\pm 0.5$& $(2592.25\pm 0.28, 2.6\pm 0.6)$ & 2630 & $2518\pm 0.5$ &$(2628.11\pm 0.19, < 0.97)$\\  \hline
 strange & 1526 & $1193\pm 4$ & $\left(1421^{+3}_{-2}{}^{+16}_{-20},38^{+16}_{-10}\pm 20\right)$ & 1537 & $1385\pm 2$ &  $(1519\pm 1$, $16\pm 1)$\\
             & & & $\left(1388\pm 9{}^{+11}_{-78}, 230\pm 50{}^{+0}_{-140}\right)$ &$-$&$-$& $-$ \\\hline\hline
    \end{tabular}
    }
    \caption{Masses and widths of the lowest-lying $J^P= 1/2^-$ and $J^P= 3/2^-$ isoscalar resonances, and numerical input for the BSE kernel-interaction and loop-function used in this work. In addition, for the pion mass and the decay constant, we take $m_\pi=139.57$ MeV and $f_\pi=92.4$ MeV, respectively. In the strange sector,  we also use $m_{\olsi{K}}=495.6$ MeV and $m_N=938.92$ MeV,  for the masses of anti-kaon and nucleon particles. For the two theoretical resonances associated to the strange $\Lambda(1405)$, we quote masses and widths and uncertainties as in Table~\ref{tab:properties}. Note that the quantum numbers of the $\Sigma_{(Q)}$ and $\Sigma_{(Q)}^{*}$ are $J^P=1/2^+$ and $3/2^+$, respectively.}
\end{table}

We will distinguish between the different charge channels when computing the decays of the bottom and charm resonances into $\Lambda_{c,b}\pi\pi$, and shall use the appropriate masses for each decay configuration.

\subsection{Chiral WT and intermediate CQM-state exchange potentials}
The S-wave chiral tree level amplitudes for isospin zero and total angular momentum $J$ read\footnote{We use here the normalizations and conventions of Ref.~\cite{Nieves:2019nol}.} 
\begin{equation}
\label{eq:WT}
[V_{\chi}]_{ij}^J(s) = D_{ij}^J \frac{2 \sqrt{s}-M_i-M_j}{4 f_\pi^2} \sqrt{\frac{E_i+M_i}{2 M_i}} \sqrt{\frac{E_j+M_j}{2M_j}},
\end{equation}
with $s$ the baryon-meson Mandelstam variable,  $M_i$ and $m_i$ the masses of the baryon and meson in the $i$ channel, respectively, and $E_i =(s-m_i^2+M_i^2)/2 \sqrt{s}$, the center-of-mass energy of the baryon in the same channel.  The $D_{ij}^J$ matrices are determined by the underlying WT structure of the interaction, with $f_\pi$ the pion decay constant.

From the discussion in the Introduction of the relevant thresholds (see Fig.~\ref{fig:espectro}), we consider
\begin{itemize}
    \item only one channel, either the $\Sigma_Q \pi$ or $\Sigma_Q^*\pi$, both in the bottom and charm sectors for $J=1/2$ or $3/2$ respectively, with $D^{J=1/2}=D^{J=3/2}=-4$ (see \textit{e.g.} Ref.~\cite{Lu:2014ina}).
    \item only the $\Sigma^*\pi$ channel in the strange $J=3/2$ sector, with also $D^{J=3/2}=-4$ (see \textit{e.g.} Ref.~\cite{Kolomeitsev:2003kt}).
    \item both the $\Sigma\pi$ and $N\olsi{K}$ coupled channels in the strange $J=1/2$ sector, with (see \textit{e.g.} Ref.~\cite{Garcia-Recio:2002yxy}):
\begin{eqnarray}
&& \begin{array}{cc} ~~~~~\Sigma\pi&   ~~~~~N\olsi{K} \end{array}\nonumber \\
D^{J=1/2} &=& \left( \begin{array}{cc}   -4  &  ~~\sqrt{3/2}   \\\\
  \sqrt{3/2} &    -3  
 \end{array}\right  ). \label{eq:DS12}
\end{eqnarray}
\end{itemize}

\begin{figure}[t]
\begin{center}
\includegraphics[height=3cm]{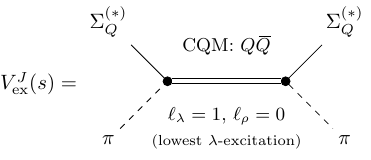}
\end{center}
\caption{Diagrammatic representation of the interaction of Eq.~\eqref{eq:exchange-0}.}\label{fig:exch}
\end{figure}

On the other hand, bare CQM-states effects on the baryon-meson dynamics can be effectively considered by means of an energy-dependent interaction~\cite{Cincioglu:2016fkm,Albaladejo:2016ztm,Albaladejo:2018mhb}, which reads:
\begin{itemize}
    \item for the heavy-quark $J=1/2$ and $3/2$ and the strange $J=3/2$ sectors (see Fig.~\ref{fig:exch}).
    \begin{eqnarray}
    V_\text{ex}^J(s)=2\Mbare\frac{d^2_{b,c,s}}{s-(\Mbare)^2}, \label{eq:exchange-0} 
    \end{eqnarray}

Note that for massless pions the $d_Q$ LEC should vanish to preserve CS. 

    \item for the two-channel strange $J=1/2$ sector
    \begin{eqnarray}
 \label{eq:exchange}
&& \begin{array}{cc} ~~~~~~\Sigma\pi&   ~~~~~~~~~N\olsi{K} \end{array}\nonumber \\
 V_\text{ex}^{J=1/2}(s)=\frac{2\Mbare}{s-(\Mbare)^2}D^{J=1/2}_{\rm CQM}\,,\qquad  D^{J=1/2}_{\rm CQM} &=&\left( \begin{array}{cc}  d_s^{\prime 2}  &  ~~\sqrt{3}d_s^\prime c_s/2   \\\\
  \sqrt{3}d_s^\prime c_s/2 &    3c_s^2/4 
 \end{array}\right  ),\label{eq:leccs}
\end{eqnarray}
\end{itemize}
where $d_{b,c,s}$, $d_s^\prime$ and $c_s$ are  undetermined dimensionless parameters  that control the strength of the baryon-meson-CQM-state vertex. The above interactions account for the contribution to the baryon-Goldstone boson scattering of the exchange of an intermediate odd-parity CQM $\lambda-$mode $j_q^\pi=1^-$ state, with masses obtained in Ref.~\cite{Yoshida:2015tia} and compiled in Table~\ref{tab:numval}. The resulting interactions are consistent with HQSS and HQFS in the $j_q^\pi=1^-$subspace of the $J^P=1/2^-$ and $J^P=3/2^-$ sectors, which are related by a spin-flavor rotation of the heavy quark. Actually, it is only thanks to these heavy-quark symmetries that the $J=1/2$ and $J=3/2$ charm and bottom sectors can be described by the same $d_b=d_c \equiv d_Q$ LEC, up to ${\cal O}(\Lambda_{\rm QCD}/m_b, \Lambda_{\rm QCD}/m_c)$ corrections~\cite{Nieves:2019nol}. Since one should expect sizable breaking of symmetry under strange-quark spin rotations, the LECs $d^\prime_s$ and $d_s$  for $J=1/2$ and $J=3/2$, respectively, are in principle unrelated, and one should also expect them to deviate from the common LEC ($\equiv d_Q$) introduced in the heavy-quark sectors. In addition, $c_s$ is introduced to account for the coupling of the $\olsi{K} N$ pair to strange the CQM $\lambda-$mode $j_q^\pi=1^-$ bare state.

Note that $ V_\text{ex}^J(s)$ introduces a pole in the baryon-meson tree-level amplitudes located at the bare mass value, $\sqrt{s}=\Mbare$. It should be interpreted as the mass of the CQM state in the limit of vanishing coupling to the baryon--meson-pairs ($d_{b,c,s},d^\prime_s,c_s\to 0$) and therefore it is not an observable. The interaction with the baryon-meson cloud \textit{dresses} the CQM state through loops, renormalizing its mass, and the dressed state might also acquire a finite width when it is located above the threshold. A priori, $\Mbare$ is a free parameter of the present approach, and moreover, it depends on the RS~\cite{Cincioglu:2016fkm}. This is because, in the effective theory, the UV regulator is finite, and the difference between the bare and the physical resonance masses is a finite renormalization that depends on the scheme.  The value of the bare mass, which is thus a free parameter, can either be indirectly fitted to experimental observations, or estimated from schemes that ignore the coupling to baryon-meson pairs, such as the CQM study carried out in Ref.~\cite{Yoshida:2015tia}. In this latter case, employed in this work,  the issue certainly would be to set the UV regulator to match the quark model and the baryon-meson scattering approaches~\cite{Cincioglu:2016fkm}.  We will explore different values of the UV cutoff and the RS-dependent couplings $d_Q$ and $d_s^{(\prime)},c_s$ to analyze the robustness of our results.
\subsection{Renormalization of the Bethe--Salpeter equation}
\label{sec:renor}
\begin{figure}[t]
\begin{center}
\includegraphics[width=0.99\textwidth]{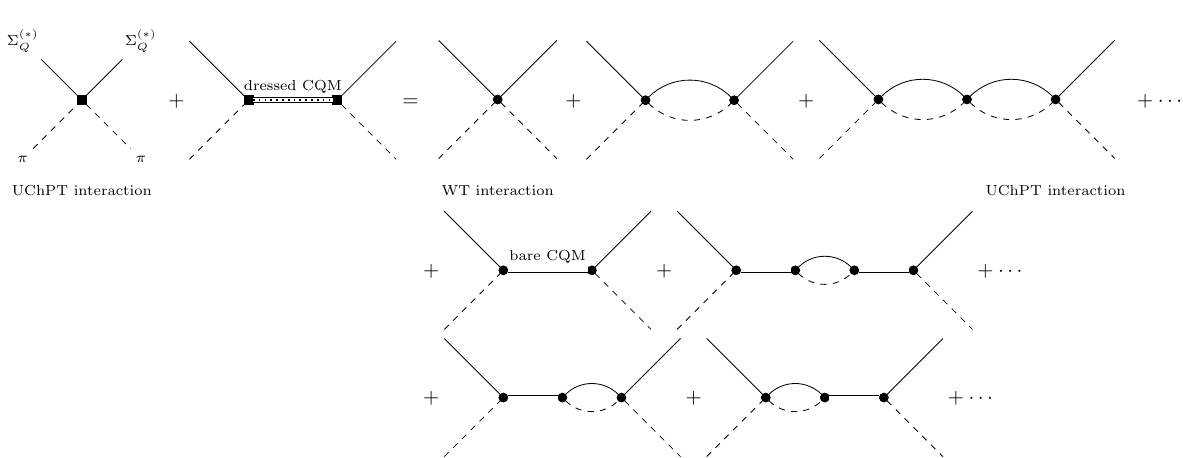}
\end{center}
\caption{Diagrammatic representation of Eq.~\eqref{eq:LS} for the single channel $\Sigma_Q^{(*)}\pi$ case, whose amplitude is given by the UChPT term, driven by the LO WT Lagrangian, plus the exchange of a dressed CQM state. }\label{fig:bse}
\end{figure}

We use the effective potential $V^J= V_\chi^J+V_\text{ex}^J$ as kernel to solve the BSE, leading to a $T$-matrix (see Fig.~\ref{fig:bse})
\begin{equation}
\label{eq:LS}
T^J(s)=\frac{1}{1-V^J(s) G^J(s)} V^J(s),
\end{equation}
satisfying exact unitarity in coupled channels in the ($\Sigma\pi,~N\overline{K}$) strange $J=1/2$ sector or elastic unitarity in the rest of single-channel sectors. The diagonal matrix $G^J(s)$ is constructed out of the baryon-meson loop functions
\begin{equation}
\label{eq:normloop}
G_i(s)=2M_i \,\, i\!\!  \int \!\! \frac{d^4 q}{(2 \pi)^4} \frac{1}{q^2-m_i^2+i\epsilon} \frac{1}{(P-q)^2-M_i^2+i\epsilon},
\end{equation}
with $P$ the total momentum of the system such that $P^2=s$. We omit
the index $J$ from here on for simplicity. The bare loop function is
logarithmically UV divergent and needs to be
renormalized. This can be done by one-subtraction
\begin{equation}
G_i(s)=\olsi{G}_i(s)+G_i(s_{i+}) ,
\label{eq:div}
\end{equation}
where the finite part of the loop function, $\olsi{G}_i(s)$, reads
\begin{equation}
\olsi{G}_i(s) = \frac{2 M_i}{(4 \pi)^2} \left(\left[ \frac{M_i^2-m_i^2}{s}-\frac{M_i-m_i}{M_i+m_i}\right] \log \frac{M_i}{m_i}+L_i(s) \right),\label{eq:defG}
\end{equation}
with $s_{i\pm}=(m_i\pm M_i)^2$, and the multi-valued function $L(s)$ given in Eq.~(A10) of Ref.~\cite{Nieves:2001wt}.

The divergent contribution of the loop function, $G_i(s_{i+})$ in Eq.~\eqref{eq:div} needs to be renormalized.  We make finite the UV divergent part of the loop function using a cutoff regulator $\Lambda$ in momentum space, which leads to~\cite{Garcia-Recio:2010enl}
\begin{eqnarray}
G_i^\Lambda(s_{i+}) &=& \frac{1}{4\pi^2} \frac{M_i}{m_i+M_i} \left
(m_i\ln\frac{m_i}{\Lambda + \sqrt{\Lambda^2+m_i^2}}+  M_i\ln\frac{M_i}{\Lambda + \sqrt{\Lambda^2+M_i^2}} \right)\label{eq:uvcut} ,
\end{eqnarray}
This RS is consistent with HQSS~\cite{Lu:2014ina} in the heavy-quark sectors and prevents from abnormally large influence of high energy channels. Note also that,  there are no cutoff effects in the  finite $\olsi{G}_i(s)-$loop function,  as it would happen if the two-body propagator of Eq.~\eqref{eq:normloop} would have been directly calculated using the UV cutoff $\Lambda$. The advantage of this RS is thus that, while it preserves the right analytical properties of the loop function, at the same time it provides some control on the subtraction constant by means of a cutoff.

\subsection%
          [Riemann sheets, poles, residues and molecular content of $\Sigma_Q^{(*)}\pi$ bound states]
{\boldmath Riemann sheets, poles, residues and molecular content of $\Sigma_Q^{(*)}\pi$ bound states }
\label{sec:observables}
Masses and widths of the dynamically generated resonances in each sector are determined from the positions of the poles, $\sqrt{s_R}$, in the SRS of the corresponding baryon-meson scattering amplitudes, namely $\sqrt{s_R}= M_R-i\, \Gamma_R/2$. In some cases, we also find real poles in the FRS which correspond to bound states. The different Riemann sheets are defined in terms of the multi-valued function $L(s)$, introduced in Eq.~\eqref{eq:defG}, that is evaluated as explained in Eq.~(A13) of Ref.~\cite{Nieves:2001wt}, and thus are labeled by $\xi_1\xi_2$ or  $\xi_1$, with $\xi_i=0,1$, in the two-channel strange $J=1/2$ sector or the single-channel sectors, respectively. The SRS in the relevant fourth quadrant is obtained from the first quadrant FRS by continuity across the corresponding unitarity cut.

The coupling constants of each bound state/resonance to the various baryon-meson states are obtained from the residues at the pole by matching the amplitudes to the expression
\begin{equation}
\left. T^J_{ij}(s)\right|_{\rm FRS/SRS}=\frac{g_i  g_j }{\sqrt{s}-\sqrt{s_R}} \ , \label{eq:pole}
\end{equation}
%
%
for values of the Mandelstam variable $s$ close to the pole, where the dimensionless couplings, $g_i$, turn out to be complex in general.

For an S-wave bound  state below the $\Sigma_Q^{(*)}\pi$ threshold,  its molecular probability can be defined through the Weinberg's compositeness rule~\cite{Weinberg:1962hj, Weinberg:1965zz}, generalized for the BSE formalism in Refs.~\cite{Gamermann:2009uq,Garcia-Recio:2015jsa},
\begin{equation}
 P_{\Sigma_Q^{(*)}\pi}=-g^2_{\Sigma_Q^{(*)}\pi}\, \left. \frac{\partial \olsi{G}_{\Sigma_Q^{(*)}\pi}(\sqrt{s})}{\partial\sqrt{s}}\right|_{\sqrt{s}=\sqrt{s_R}}\label{eq:prob}
\end{equation}
where the loop function is evaluated in the FRS. There is a relationship between the molecular content of a two-hadron S-wave bound state and the scattering length and effective range parameters. We refer the reader to recent works~\cite{Albaladejo:2022sux,Song:2022yvz,Kinugawa:2022fzn} on the subject. In addition, we have 
\begin{equation}
    \frac{1}{g^2_{\Sigma_Q^{(*)}\pi}} = \frac{\partial}{\partial\sqrt{s}} \left[V^{-1}_{\Sigma_Q^{(*)}\pi}(\sqrt{s})-\olsi{G}_{\Sigma_Q^{(*)}\pi}(\sqrt{s})\right]_{\sqrt{s}=\sqrt{s_R}} \label{eq:gcoup}
\end{equation}
where $V_{\Sigma_Q \pi}$ and $V_{\Sigma_Q^*\pi}$ are the full single-channel potentials $V^J= V_\chi^J+V_\text{ex}^J$ with $J=1/2$ and $3/2$, respectively. While the dependence of $g^2_{\Sigma_Q^{(*)}\pi}$ on $d_Q$ is explicit, through  $V_\text{ex}$, that on the UV cutoff is implicit (hidden) in the dependence of the pole-position $s_R$ on both $d_Q$ and $\Lambda$. 
\vspace{0.25cm}

\begin{figure}[t]
\begin{center}
\includegraphics[width=0.8\textwidth]{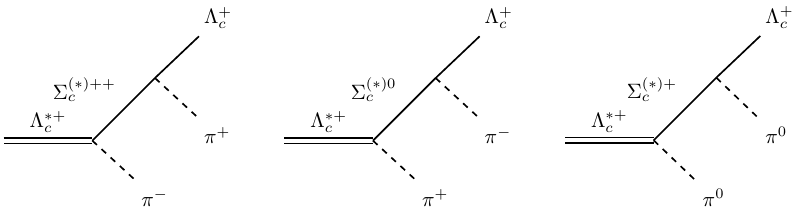}
\end{center}
\caption{Diagrams for $\Lambda_c(2595)$ or $\Lambda_c(2625)$ decay into $\Lambda_c^+$ and two pions, mediated by the $\Sigma_c$ or $\Sigma^*_c$ resonances, respectively.\label{fig:decay}}
\end{figure}
Finally, the coupling  $g_{\Sigma_Q^{(*)}\pi}$ can also be used to evaluate the $\Lambda^*_Q(1/2^-,3/2^-)\to \Lambda_Q(1/2^+)\pi\pi$ decay through the $\pi\Sigma^{(*)}_Q$ intermediate state (see Fig.~\ref{fig:decay} for the $\Lambda_c(2595)$ or $\Lambda_c(2625)$ decays), which is the only open channel in the bottom and charm sectors. These resonant contributions to the three-body decay widths, $\Gamma_{\Lambda_Q\pi\pi}^R$ will be calculated in this work as explained in detail in the Subsec. IIE of Ref.~\cite{Nieves:2019nol}, where explicit expressions for the charm-sector are given. As mentioned above,  we use the appropriate (physical) masses for each decay configuration because in some cases the available phase-space is quite limited and it is very sensitive to the masses of the involved hadrons.

\section{Spectrum}
\label{sec:spectrum}
\subsection%
          [Determination of $d_Q(\Lambda)$ and FRS amplitudes in the bottom sector: The $\Lambda_b(5912)$ and $\Lambda_b(5920)$ resonances]
{\boldmath Determination of $d_Q(\Lambda)$ and FRS amplitudes in the bottom sector: The $\Lambda_b(5912)$ and $\Lambda_b(5920)$ resonances}
\label{sec:dq-bottom}
\begin{figure}[t]
\begin{center}
\makebox[0pt]{\includegraphics[width=0.5\textwidth]{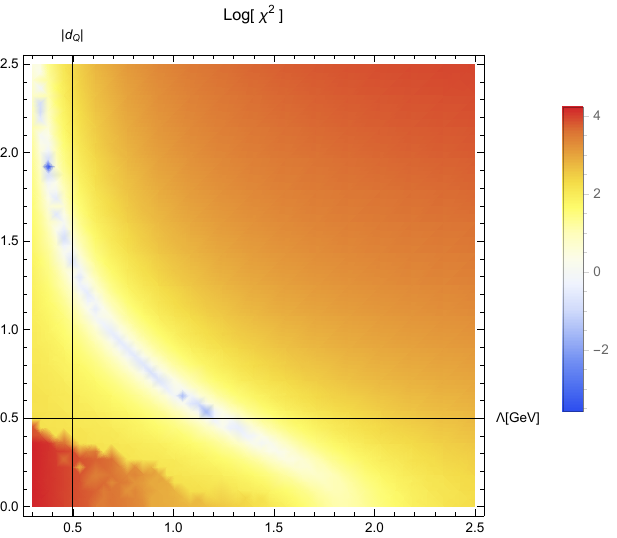}\hspace{0.5cm}\includegraphics[width=0.5\textwidth]{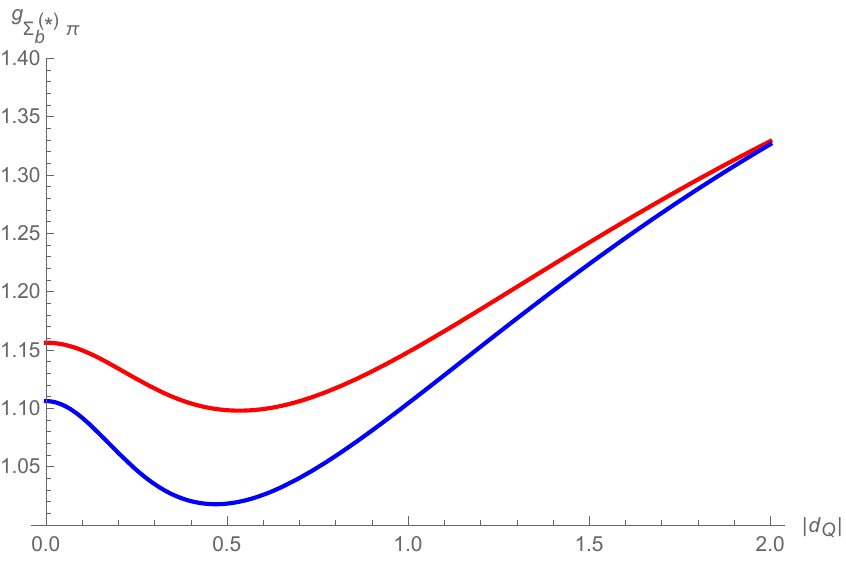}}\\\vspace{0.5cm}
\makebox[0pt]{\includegraphics[width=0.5\textwidth]{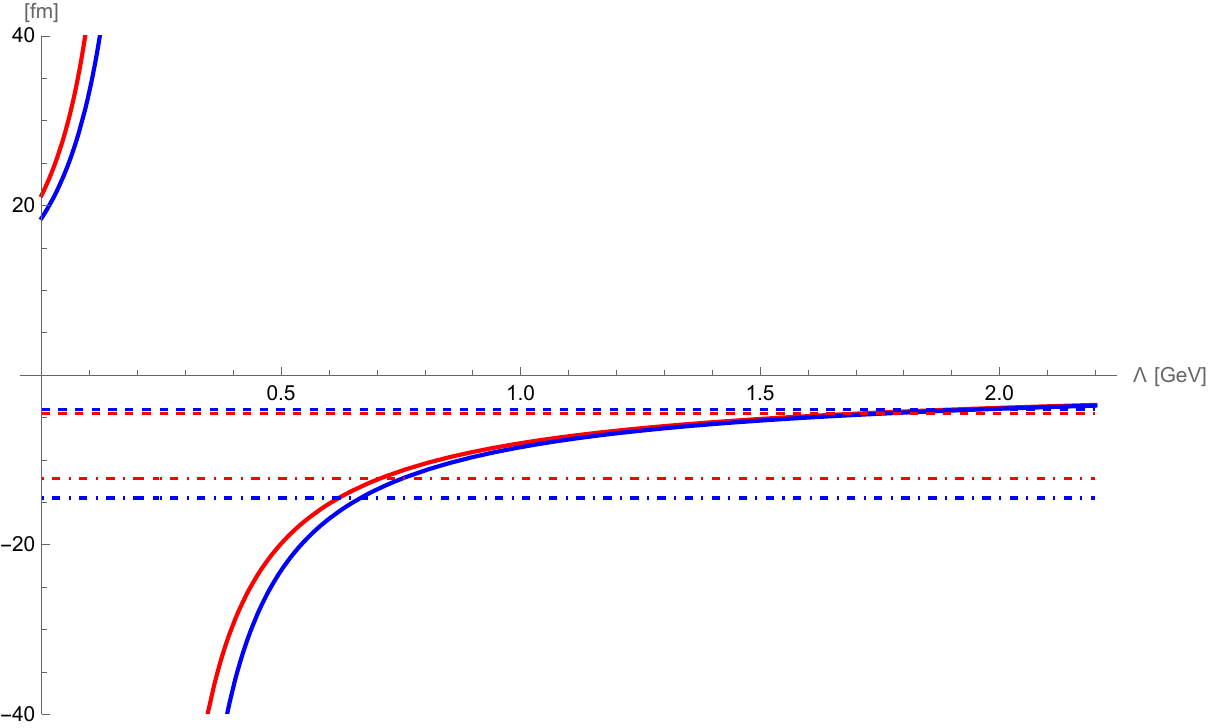}}
\end{center}
\caption{Top left: Density plot of the decadic logarithm of the merit function $\chi^2$ [Eq.~\eqref{eq:chi2}] as a function of the UV cutoff $\Lambda$ (x-axis) and the coupling $|d_Q|$ (y-axis), which controls the interplay between the baryon-meson and CQM degrees of freedom. Top right: Dependence of the couplings $g_{\Sigma_Q \pi}$ (red) and $g_{\Sigma_Q^*\pi}$ (blue) on $|d_Q|$ as obtained from Eq.~\eqref{eq:gcoup} setting in each case the pole position $\sqrt{s_R}$  to the $\Lambda_b(5912)$ and $\Lambda_b(5920)$ resonance masses. Bottom: Solid curves show the UV cutoff dependence of the FRS inverse loop function, in fermi units, for both $J=1/2$ (red) and $J=3/2$ (blue)  bottom channels evaluated at the  $\Lambda_b(5912)$ and $\Lambda_b(5920)$ physical masses, respectively. Dashed and dash-dotted lines show the chiral interaction ($V_\chi$) and total effective potential ($V=V_\chi+V_{\rm ex}[\,|d_Q|=1]$), in fermi units respectively, evaluated for the nominal masses of the resonances in the $J=1/2$ (red) and $J=3/2$ (blue) bottom sectors.}\label{fig:V}
\end{figure}

To begin with, we consider the heavy-quark sectors and only one free parameter ($d_Q$) in the BSE interaction kernel, which will depend on the UV cutoff. We determine $d_Q$ by fitting the masses of the $\Lambda_b(5912)$ and $\Lambda_b(5920)$ resonances, which within the present scheme appear as bound states. HQSS breaking corrections in the bottom sector are expected to be small $[{\cal O}(\Lambda_{\rm QCD}/m_b)\sim 5\%]$, and therefore using the same LEC $d_Q$ to simultaneously describe the  $J=1/2$ and $J=3/2$ sectors should be accurate enough.  Furthermore, the numerical value extracted from the bottomed baryons for this LEC should be also a good approximation to its value in the infinite mass limit. On the other hand, as can be seen in Fig.~\ref{fig:espectro}, neither the $\Sigma_b^{(*)}\pi$ thresholds nor the CQM states are located too close the $\Lambda_b(5912)$ and $\Lambda_b(5920)$, as it happens for instance in the charm sector. Thus the masses of these resonances are expected to be rather sensitive to both meson-baryon and CQM degrees of freedom. We minimize the $\chi^2$-function  defined as
\begin{eqnarray}
 \chi^2(|d_Q|, \Lambda) = \left[\frac{M_{\Lambda_b(5912)}-M_{R-{\rm BSE}}^{J^P=1/2^-}(|d_Q|, \Lambda)}{\sigma(\Sigma_b)}\right]^2  + \left[ \frac{M_{\Lambda_b(5920)}-M_{R-{\rm BSE}}^{J^P=3/2^-}(|d_Q|, \Lambda)}{\sigma(\Sigma_b^*)}\right]^2  \label{eq:chi2}
\end{eqnarray}
with $M_{R-{\rm BSE}}^{J^P}$ the position below the threshold of the pole on the real axis of the FRS of the theoretical amplitude. To define the merit function, we have used the errors quoted in Table~\ref{tab:numval} for the charged averaged masses of the $\Sigma_b^{(*)}$ baryons instead of those smaller given in the RPP for the $\Lambda_b(5912)$ and  $\Lambda_b(5920)$ masses.
In this way, we take into account the systematic uncertainty affecting our approach due to isospin breaking.  

In the top left plot of Fig.~\ref{fig:V}, we display the base-10 logarithm of $\chi^2$ as a function of $|d_Q|$ and $\Lambda$. The UV cutoff is varied from $0.3\,\text{GeV}$ to $2.5\,\text{GeV}$.\footnote{Note that the real parts of the FRS $\Sigma_b\pi$ and $\Sigma_b^*\pi$ loop functions evaluated at the physical $\Lambda_b(5912)$ and $\Lambda_b(5920)$ physical masses, respectively, are negative for $\Lambda \gtrsim 0.3\,\text{GeV}$. For smaller values of the cutoff, around $0.21\,\text{GeV}$ for $G_{\Sigma_b\pi}(s=M_{\Lambda_b(5912)}^2)$ and $0.25\,\text{GeV}$ for $G_{\Sigma^*_b\pi}(s=M_{\Lambda_b(5920)}^2)$, the loop functions in the FRS pass through zero and change sign, becoming positive, which eventually allow the appearance of bound states with a repulsive potential. Thus, $\Lambda=0.3\,\text{GeV}$ is a reasonable lower limit to explore the dependence of $\chi^2(|d_Q|, \Lambda)$ on the UV cutoff.} It can be seen that values of $\chi^2 \leqslant 1$ can be obtained only for cutoffs below $1.4$--$1.5\,\text{GeV}$, appropriately choosing the LEC $|d_Q|$ for each $\Lambda$. For the latter, we consider reasonable to let it vary between $0.4$--$1.2\,\text{GeV}$.  Results are compiled in Table~\ref{tab:bottom-res}.
\begin{table}[t] 
\centering
\resizebox{1.\textwidth}{!}{
\begin{tabular}{c|cc|cccc|cccc}
\hline\hline
   \multicolumn{1}{c|}{} &  \multicolumn{2}{c|}{} & \multicolumn{4}{c|}{$\Lambda_b(5912)$}   &\multicolumn{4}{c}{ $\Lambda_b(5920)$}    \\
\hline 
 \multicolumn{1}{c|}{} &  \multicolumn{2}{c|}{}  & \multicolumn{4}{c|}{$\Sigma_b\pi$ $J^P={\frac{1}{2}}^-$}   & \multicolumn{4}{c}{$\Sigma^*_b\pi$ $J^P={\frac{3}{2}}^-$}     \\
 \hline
 $\Lambda$ [GeV]  & $\chi^2$ & $|d_Q|$  & $M$ [MeV] & $|g_{\Sigma_b\pi}|$ &$P_{\Sigma_b\pi}$ & $\Gamma_{\Lambda_b\pi\pi}^R$ [keV] &  $M$ [MeV] & $|g_{\Sigma_b^*\pi}|$ &$P_{\Sigma^*_b\pi}$ & $\Gamma_{\Lambda_b\pi\pi}^R$ [keV]  \\
 \hline
$0.4$   & 0.02 &  $1.79\pm 0.11$      &    $5912.4\pm 2.0$ & $1.67\pm 0.06$  &  $0.35\pm 0.02$ & $18 \pm 5$ & $5919.8\pm 1.6$ & $1.66\pm 0.07$  &$0.31\pm 0.03$  &  $37 \pm 5$  \\
$0.65$  & 0.32 &  $1.06\pm 0.06$      &    $5913.1\pm 2.0$ & $1.34\pm 0.04$  &  $0.23\pm 0.01$ & $13 \pm 4$ & $5919.1\pm 1.7$ & $1.26\pm 0.05$  &$0.18\pm 0.01$  &  $19 \pm 3$ \\
$0.9$  & 0.16 &  $0.75\pm 0.04$      &    $5912.9\pm 1.7$ & $1.23\pm 0.03$  &  $0.19\pm 0.01$ & $10 \pm 3$ & $5919.5\pm 1.6$ & $1.11\pm 0.04$  &$0.14\pm 0.01$  &  $16 \pm 3$   \\
$1.15$    & 0.00 &  $0.55\pm 0.04$      &    $5912.1\pm 2.0$ & $1.21\pm 0.02$  &  $0.18\pm 0.01$ & $9 \pm 3$ & $5920.2\pm 1.9$ & $1.04\pm 0.03$  &$0.12\pm 0.01$  &  $15 \pm 3$    \\
\hline
$1.85\pm 0.04$    & 12 &  0      &    $5905.5\pm 1.7$ & $1.27\pm 0.02$  &  $0.19\pm 0.01$ & $2.5 \pm 1.2$ & $5924.9\pm 1.7$ & $1.27\pm 0.02$  &$0.19\pm 0.01$  &  $39 \pm 8$ \\
\hline
\end{tabular}
}
\caption{Fit results and masses of the $\Lambda_b(5912)$ and $\Lambda_b(5920)$ resonances for different UV cutoffs. For each case, we also give the associated predictions for the $g_{\Sigma_b^{(*)}\pi}$ couplings, $\Sigma_b^{(*)}\pi$ molecular probabilities and the three-body decay $\Gamma_{\Lambda_b\pi\pi}^R$ widths, discussed in Subsec.~\ref{sec:observables}. Errors on $|d_Q|$ are obtained as the shift around the best-fit value, such that $\chi^2$ increases by one unit. These errors are transported to the rest of the properties collected in the table. In the last row, we show the results obtained when $d_Q$ is set to zero and the UV cutoff is fitted. }\label{tab:bottom-res}
\end{table}

We see that the two states $\Lambda_b(5912)$ and $\Lambda_b(5920)$ can be described simultaneously very well for the entire range of UV cutoffs considered here, which shows that HQSS performs quite well in the bottom sector. As $\Lambda$ increases, the parameter $|d_Q|$ (which controls the interplay between hadron and CQM degrees of freedom) decreases, and also the $g_{\Sigma_Q \pi}$ and $g_{\Sigma_Q^*\pi}$ couplings, as inferred from the curves depicted in the top right panel of Fig.~\ref{fig:V}.\footnote{ Note that in this plot, the positions $\sqrt{s_R}$ are fixed to the physical masses of the resonances $\Lambda_b(5912)$ and $\Lambda_b(5920)$, which implies that for a given value of $|d_Q|$, the UV cutoff should be separately tuned in each $J^P$ sector. However, the curves displayed in the plot describe quite accurately the couplings collected in Table~\ref{tab:bottom-res}, obtained from a common cutoff and fitting $|d_Q|$,  since $\chi^2$ is always small. The couplings collected in the last row of the table are much worse reproduced since in that case the fitted theoretical pole masses significantly differ from the physical ones.}

The $g_{\Sigma_Q^{(*)}\pi}$ couplings determine the molecular probabilities of the bound states [\textit{cf.} Eq.~\eqref{eq:prob}], which turn out to be small in the range $0.15$--$0.35$, with  $\left.\partial \olsi{G}_{\Sigma_Q^{(*)}\pi}/\partial\sqrt{s}\right|_{\sqrt{s}=\sqrt{s_R}}\sim -0.1$. Thus, both the $\Lambda_b(5912)$ and $\Lambda_b(5920)$ are largely CQM states, though the chiral $\Sigma_b^{(*)}\pi$ interaction becomes essential to accurately describe their masses. This is the first important result of this work. We also emphasize that setting the CQM-state exchange potential ($V_{\rm ex}$) to zero (bottom row of Table~\ref{tab:bottom-res}; see also top left panel of Fig.~\ref{fig:V}) leads to a much poorer description of the $\Lambda_b(5912)$ and $\Lambda_b(5920)$ masses. Furthermore, such a scheme requires large UV cutoffs of the order of $1.8$--$1.9\,\text{GeV}$, which, however, are not capable of satisfactorily accounting for effects produced by the $V_{\rm ex}$ interaction.

In addition, the three-body decay $\Gamma_{\Lambda_b\pi\pi}^R$ widths, mediated by the intermediate $\Sigma_b^{(*)}$ states (see Fig.~\ref{fig:decay}), are very small, of the order of tens of keV and can be easily accommodated in the lower bounds of hundreds of keV for the $\Lambda_b(5912)$ and $\Lambda_b(5920)$ widths reported in the RPP~\cite{ParticleDataGroup:2022pth}. We predict small $\Gamma_{\Lambda_b\pi\pi}^R$ widths since, as can be seen in Fig.~\ref{fig:espectro}, the intermediate $\Sigma_b^{(*)}$ are quite off-shell in the decay, as compared to their widths of the order of $5$--$10\,\text{MeV}$. Hence this mechanism might not be the dominant one in the decay of the $\Lambda_b(5912)$ and $\Lambda_b(5920)$ into $\Lambda_b\pi\pi$, while it could be more important in the charm sector, in particular for the $\Lambda_c(2595)\to \Lambda_c\pi\pi$ reaction, since it is very sensitive to the available phase-space.

We pay now attention to the bottom panel of Fig.~\ref{fig:V}, which can help us to understand the UV cutoff dependence of $d_Q$ seen in Table~\ref{tab:bottom-res}. Bound states manifest as poles ($s_R$) in the FRS of the $T$-matrix. They are located in the real axis below the $\Sigma_b^{(*)}\pi$ threshold and satisfy $V(s_R)=1/G(s_R)$. In the plot, we show $1/G(s_R)$ in the FRS as a function of $\Lambda$. The FRS loop function evaluated at the mass of the $\Lambda_b(5192)$ or the $\Lambda_b(5920)$, depending on the channel, has a zero around $\Lambda\sim 0.2$ GeV, and below this cutoff becomes positive. Thus, the pole condition should be satisfied by more and more attractive potentials as  $\Lambda$ decreases and approaches from above this value around 0.2 GeV, for which $G(s_R)$ vanishes. For smaller cutoffs, there do not exist bound states. The chiral interaction $V_\chi$ is also shown in Fig.~\ref{fig:V} (dashed lines) and we see that it can generate both resonances using values of $\Lambda$ in the region of $1.8$--$1.9\,\text{GeV}$ (see last row of Table~\ref{tab:bottom-res}). This is in qualitative agreement with the UChPT results of Ref.~\cite{Lu:2014ina}, though in this latter reference even higher cutoffs of around $2.2$--$2.6\,\text{GeV}$ were required,\footnote{The difference is produced by the consideration in \cite{Lu:2014ina} of the higher-energy chiral channels $\Xi_b^{\prime}K$ for $J=1/2$ and $\Xi_b^\ast K$  for $J=3/2$, which thresholds are more than 500 MeV above the masses of the $\Lambda_b(5192)$ or $\Lambda_b(5920)$ (see Fig.~\ref{fig:espectro}). They produce rather energy-independent interactions in the region of interest that interfere with the constant part of the loop function, controlled by the UV cutoff (Eq.~\eqref{eq:uvcut}). In the $J=1/2$ sector, the {\it ldof} in the $\Lambda_b\eta$ and $\Xi_b K$ pairs, whose thresholds are lower than that of $\Xi_b^{\prime}K$, are coupled to $j_q^\pi=0^-$ and therefore these channels do no couple to $\Sigma_b\pi$ in the LO chiral WT Lagrangian, since it is consistent with HQSS. In the work of Ref.~\cite{Lu:2014ina},  different UV cutoffs are used in the $J=1/2$ ($\Lambda=2.17$ GeV) and $J=3/2$ ($\Lambda=2.60$ GeV) sectors, fine-tuned to reproduce the masses of the $\Lambda_b(5192)$ and $\Lambda_b(5920)$ states.} as mentioned in Sec.~\ref{sec:uchpt} of the Introduction. The intermediate CQM-state exchange potential $V_{\rm ex}$ is attractive and its impact on the total effective potential for $d^2_Q=1$ can be seen in Fig.~\ref{fig:V} (dashed-dotted lines). As we decrease (increase) the UV cutoff, the bound-pole condition should be satisfied by more (less) attractive total interactions, which are constructed by employing larger (smaller) values of  $d^2_Q$. For cutoffs above 1.8-1.9 GeV, the chiral interaction would need to be supplemented by a repulsive potential, which cannot be obtained from the exchange of a single CQM state, located above the physical mass of the resonance, which is always attractive.

\subsection%
          [Poles in the SRS bottom amplitudes and the $\Lambda_b(6070)$]
{\boldmath Poles in the SRS bottom amplitudes and the $\Lambda_b(6070)$}\label{sec:bottom-poles}

\begin{figure}[t]
\begin{center}
\makebox[0pt]{\includegraphics[width=0.39\textwidth]{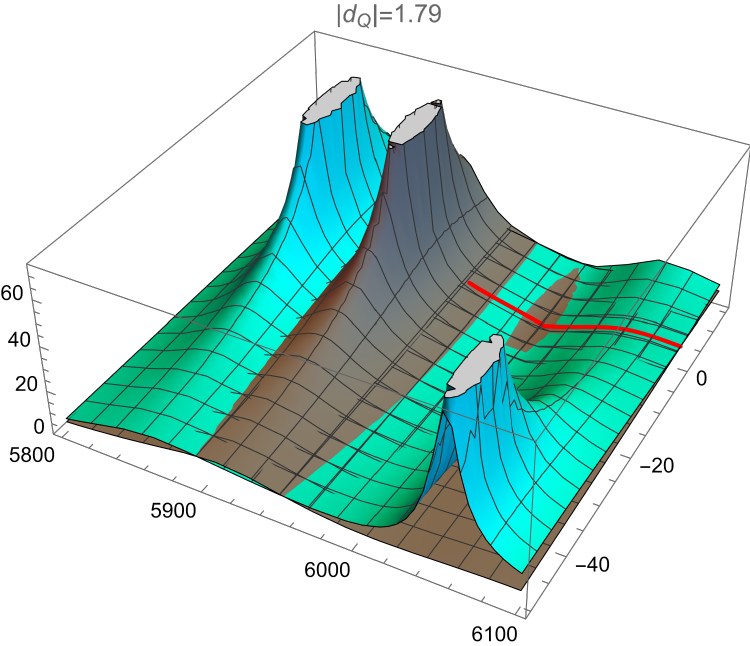}\hspace{0.75cm}\includegraphics[width=0.39\textwidth]{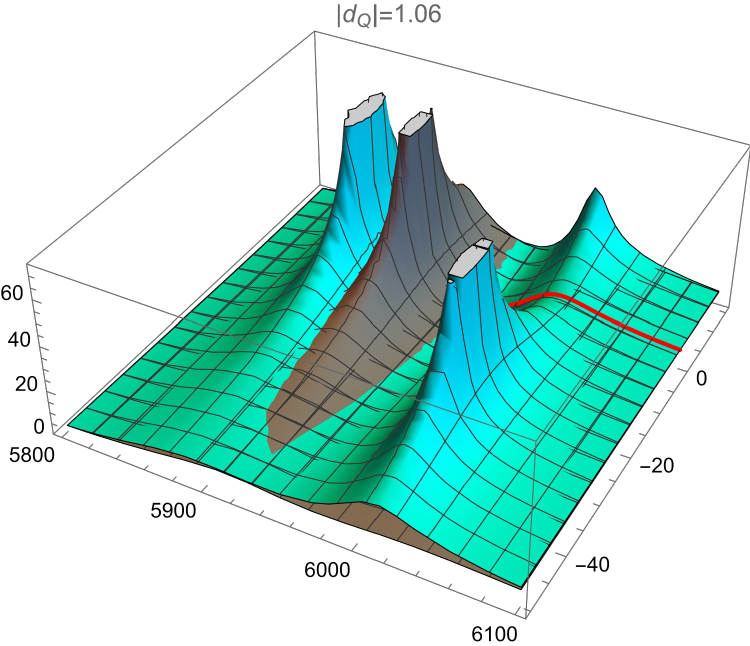}}\\\vspace{0.2cm}
\makebox[0pt]{\includegraphics[width=0.39\textwidth]{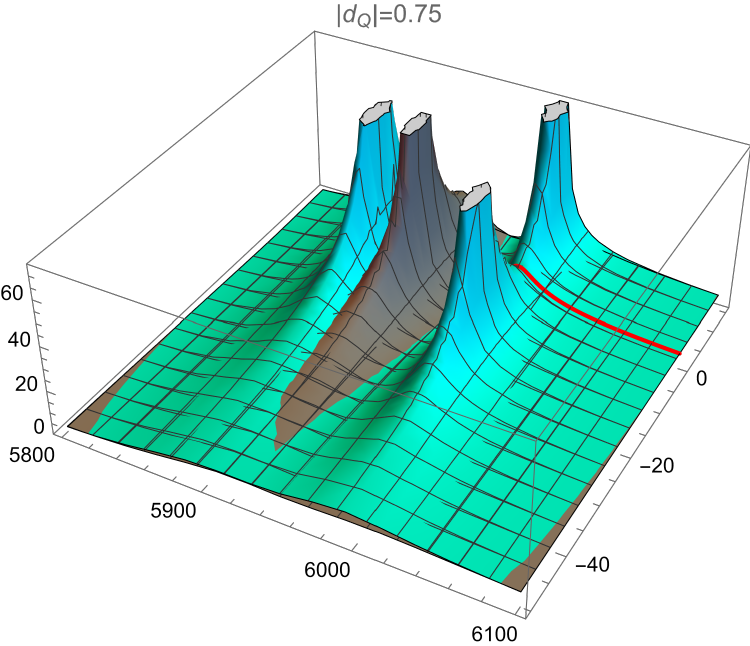}\hspace{0.75cm}\includegraphics[width=0.39\textwidth]{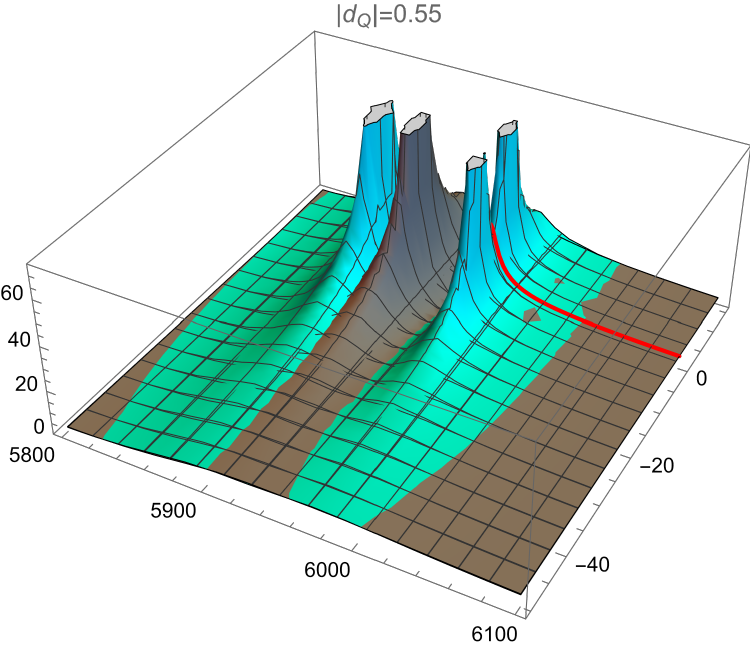}}\\\vspace{0.2cm}
\makebox[0pt]{\includegraphics[width=0.39\textwidth]{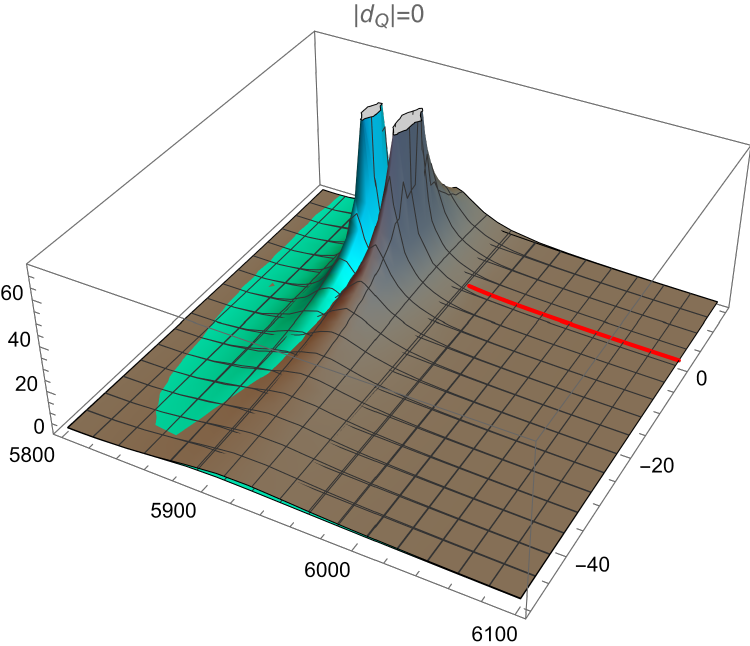}}
\end{center}
\caption{ Absolute value of the  $T_{\Sigma_b\pi}$-matrix (in fermi units), both in the first (gray) and second (greenish hues) Riemann sheets and for the values of $|d_Q|$ and $\Lambda$ considered in Table~\ref{tab:bottom-res}, as a function of complex  $\sqrt{s}=x+iy$ in MeV. We also show the scattering line (red solid line) in all the cases.}\label{fig:FRSb}
\end{figure}

In addition to the bound states discussed in the previous subsection, the unitarized amplitudes have poles in the SRS which correspond to virtual states or resonances, as can be seen in Fig.~\ref{fig:FRSb} for the $J^P=1/2^-$ sector. Results are qualitatively similar for the $T_{\Sigma_b^*\pi}$-matrix $(J^P=3/2^-)$. Virtual states (SRS poles below threshold) have little relevance since they are located behind the bound states and hence their effects are overshadowed by the physical $\Lambda_b(5192)$ and $\Lambda_b(5920)$. However, resonances (SRS poles above threshold) are QCD states which affect the $\Sigma_b^{(*)}\pi$ scattering. We see in  Fig.~\ref{fig:FRSb} that the resonance pole approaches the threshold and becomes narrower as $|d_Q|$ decreases. Its effects on the scattering line (red curve in the plots) are in all cases clearly visible and its exact position should allow for disentangling between the different scenarios contemplated in  Table~\ref{tab:bottom-res}. Note also that the present scheme does not produce any near-threshold resonance when the CQM-state exchange potential ($V_{\rm ex}$) is set to zero. In Table~\ref{tab:bottom-resII} we compile the properties of the resonances predicted by the present HQSS scheme in the bottom sector. 

\begin{table}[t] 
\centering
\resizebox{1.\textwidth}{!}{
\begin{tabular}{cc|cccc|cccc}
\hline\hline
\multicolumn{2}{c|}{} & \multicolumn{4}{c|}{$\Sigma_b\pi$ [$J^P=1/2^-$]}   & \multicolumn{4}{c}{$\Sigma^*_b\pi$ [$J^P=3/2^-$]}    \\
\hline
 $\Lambda$ [GeV]  & $|d_Q|$  & $M$ [MeV]& $\Gamma$ [MeV] &$|g_{\Sigma_b\pi}|$ & $\phi_{\Sigma_b\pi}$ &  $M$ [MeV] & $ \Gamma$ [MeV] &$|g_{\Sigma^*_b\pi}|$ & $\phi_{\Sigma^*_b\pi}$ \\
 \hline
 $0.40$   &  $1.79\pm 0.11$      &    $6053\pm 6$ & $85.2\pm 0.4$  &  $1.60\pm 0.03$ & $-0.70\pm 0.01$ &  $6066 \pm 6$ & $90.0\pm 0.5$  &$1.65\pm 0.03$ & $-0.67\pm 0.01$  \\
$0.65$  &  $1.06\pm 0.06$      &    $6008\pm 3 $ & $49.6\pm 0.5$ &  $1.46\pm 0.02$& $-0.53\pm 0.01$   & $6021 \pm 3$ & $52.9\pm 0.4$  &$1.54\pm 0.02$ & $-0.50\pm 0.01$ \\
$0.90$  &  $0.75\pm 0.04$      &    $5983\pm 3$ & $24.5\pm 0.7$  &  $1.23\pm 0.01$ & $-0.41\pm 0.01$ &  $5995 \pm 2$ & $25.9\pm 0.8$  &$1.35\pm 0.01$ & $-0.38\pm 0.01$   \\
$1.15$    &  $0.55\pm 0.04$      & $5966\pm 3$ & $9.5\pm 1.1$  &  $0.97\pm 0.01$ & $-0.30\pm 0.01$ &  $5976 \pm 3$ & $7\pm 2 $  &$1.15^{+0.06}_{-0.02}$ & $-0.30^{+0.01}_{-0.05}$      \\
 \hline\hline
\end{tabular}
}
\caption{Properties of the $\Sigma_b^{(*)}\pi$ resonances (SRS poles above threshold)  predicted in the bottom sector for the different UV cutoffs contemplated in  Table~\ref{tab:bottom-res}. In the case of the couplings, we give both the absolute value and the phase in radians. For reference, the thresholds used in this work are $(M_{\Sigma_b}+m_\pi)=5952.7\pm 2.5$ MeV and $(M_{\Sigma_b^*}+m_\pi)=5972.1\pm 2.2$ MeV.} \label{tab:bottom-resII}
\end{table}

The results presented in the table are quite robust because, as can be seen in Fig.~\ref{fig:espectro}, the closest high-energy channels discarded in the scheme would be  $N \olsi{B}$ ($J^P=1/2^-$) and $N \olsi{B}{}^*$ ($J^P=1/2^-,3/2^-$), whose thresholds are around $6220\,\text{MeV}$ and $6265\,\text{MeV}$, respectively, and thus more than $200\,\text{MeV}$ above the highest resonances in Table~\ref{tab:bottom-resII}. The chiral $j^\pi_q=0^-$ $\Omega_b\eta$ and $\Xi_b K$ channels almost do not couple to the  $j^\pi_q=1^-$ $\Sigma_b\pi$ pair because of HQSS, while the $\Xi_b^{\prime,\ast} K$ thresholds are above bar $6400\,\text{MeV}$. Finally, the $j^\pi_q=1^-$ $\rho-$mode CQM excitations are above and very close to the $N\olsi{B}{}^*$ threshold.   

\begin{figure}[t]
\begin{center}
\includegraphics[height=7cm,keepaspectratio]{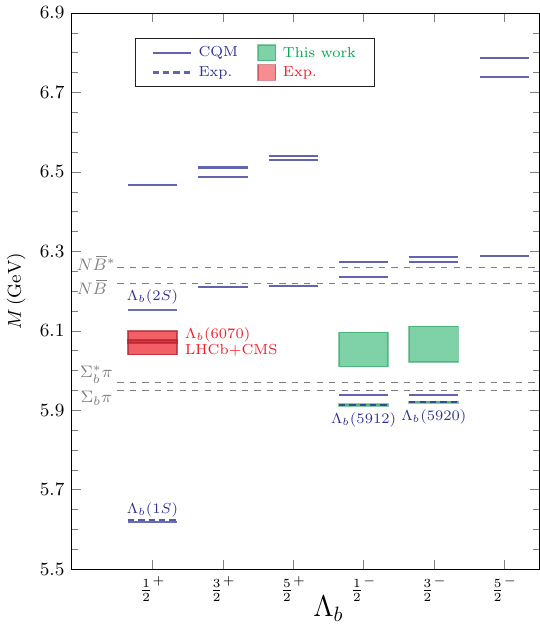}
\includegraphics[height=7cm,keepaspectratio]{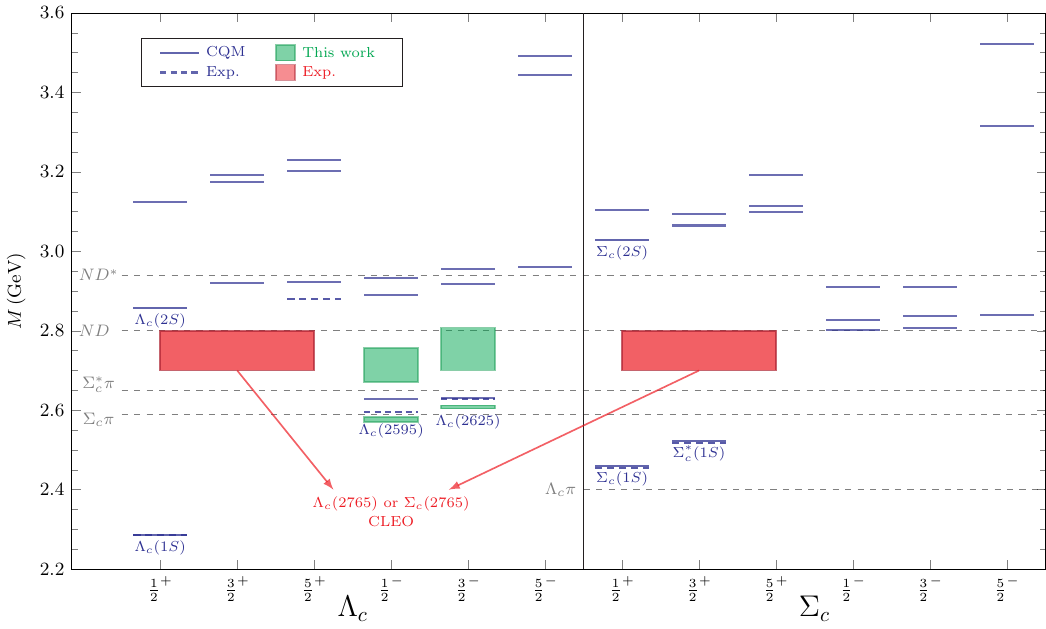}
\end{center}
\caption{Spectrum in the $\Lambda_b$ bottom (left) and the $\Lambda_c$ and $\Sigma_c$ charm (right) sectors and for different spin-parities. Some relevant thresholds are depicted by continuous dashed lines. The solid blue lines (denoted with ``CQM'') correspond to the CQM calculation of Ref.~\cite{Yoshida:2015tia}, while experimentally well-established states (``Exp.'') are depicted in dashed blue lines. In the left (right) panel, the red rectangles (also denoted by ``Exp.'') stand for the broad excess of events observed by the LHCb~\cite{LHCb:2020lzx} and CMS~\cite{CMS:2020zzv} (CLEO~\cite{CLEO:2000mbh}) in the region of $6040$--$6100\,\text{MeV}$ ($2765\,\text{MeV}$) of the $\Lambda_b^0\pi^+\pi^-$ ($\Lambda_c^+\pi^+\pi^-$) spectrum, that gives rise to the existing evidence for the $\Lambda_b(6070)$ [$\Lambda_c(2765)$], and whose spin-parity has not been experimentally determined. In the charm sector CLEO reported a broad and statistically significant peak, although nothing at its quantum numbers are not known either, including whether it is a $\Lambda_c^+$ or a $\Sigma_c$, or whether the width might be due to overlapping states. Finally, the green levels and rectangles represent the findings of our work. The lowest ones (levels) in each flavour and spin-parity sector correspond to the low-lying states in Table~\ref{tab:bottom-res} and the upper part of Table~\ref{tab:charm-res}. The higher green areas cover the range $M \pm \Gamma/2$ of the higher-energy resonances found in this work for cutoff values $\Lambda=400\,\text{MeV}$, as reported in Table~\ref{tab:bottom-resII} and the lower part of Table~\ref{tab:charm-res}.\label{fig:high}}
\end{figure}
The spectrum of excited beauty baryons decaying to the $\Lambda_b\pi^+\pi^-$ final state near threshold has been measured with good precision by the LHCb collaboration using a data sample of $pp$ collisions, collected at center-of-mass energies $7$, $8$, and $13\,\text{TeV}$, corresponding to an integrated luminosity of 9 fb$^{-1}$. In addition to the $\Lambda_b(5912)$ and $\Lambda_b(5920)$ doublet, LHCb has reported another doublet of narrow  resonances\footnote{The measured masses and widths (around 2-3 MeV) of these states are compatible with the expectations for the $3/2^+-5/2^+$ HQSS $\Lambda_b(1{\rm D})$ doublet \cite{Chen:2019ywy,Wang:2019uaj,Liang:2019aag}.} $\Lambda_b(6146)$ and $\Lambda_b(6152)$ \cite{LHCb:2019soc} and interestingly in 2020, a broad state, whose mass and width are measured to be $M=6072.3 \pm 2.9 \pm 0.6 \pm 0.2$ MeV and $\Gamma=72 \pm 11\pm 2$ MeV \cite{LHCb:2020lzx}.  From the $pp$ data at 13 TeV, the CMS collaboration \cite{CMS:2020zzv} also reported that the $\Lambda_b\pi^+\pi^-$  mass distribution exhibits a broad excess of events in the region of $6040-6100\,\text{MeV}$, corresponding to a statistical significance of four standard deviations and consistent with the new $\Lambda_b(6070)$ state claimed by LHCb.

This experimental evidence clearly supports that the appropriate UV cutoff in our model should be chosen around $400\,\text{MeV}$, which would allow us to explain the excess of events leading to the LHCb claim of the $\Lambda_b(6070)$ state as originated by the broad resonances ($\Gamma \sim 85-90\,\text{MeV}$) that we find in both $J^P=1/2^-$ and $J^P=3/2^-$ sectors with masses around $6050-6070\,\text{MeV}$.  As can be seen in Table~\ref{tab:bottom-resII}, the present model with $\Lambda \sim 1\,\text{GeV}$  predicts much narrower resonances ($\Gamma\leqslant 20\,\text{MeV}$) below $6\,\text{GeV}$,  which have not been seen in the $\Lambda_b\pi^+\pi^-$ spectrum by either LHCb or CMS collaborations, and therefore values around 1 GeV for the UV cutoff are phenomenologically disfavored. Furthermore, from a theoretical point of view, and with the idea of making predictions in the charm quark sector, it is preferable to use cutoffs smaller than $1\,\text{GeV}$, since higher values could induce considerable violations of the spin and flavor heavy quark symmetries.

The spin and parity of the $\Lambda_b(6070)$ were not established in the LHCb work~\cite{LHCb:2020lzx}, and the quantum numbers quoted in the RPP~\cite{ParticleDataGroup:2022pth} are based on the quark model expectation that the mass of the excited $\Lambda_b(2{\rm S})$ is in the range $6050--6150\,\text{MeV}$~\cite{Capstick:1986ter,Roberts:2007ni,Ebert:2011kk,Yoshida:2015tia}.  It might well be that the experimental broad signal associated to the $\Lambda_b(6070)$ will be the result of the contribution to the $\Lambda_b\pi^+\pi^-$spectrum, not only of both $J^P=1/2^-$ and  $J^P=3/2^-$ resonances found in this work, but also perhaps of some additional strength from the $J^P=1/2^+$ $\Lambda_b(2{\rm S})$ radial excitation. This is to say the width observed in the $\Lambda_b\pi^+\pi^-$spectrum might be due to overlapping states. The left panel of Fig.~\ref{fig:high} shows graphically how the resulting spectrum would look like. 

\subsection%
          [Charm sector: the $\Lambda_c(2595)$, $\Lambda_c(2625)$ and $\Lambda_c(2765)$ resonances]%
{\boldmath Charm sector: the $\Lambda_c(2595)$, $\Lambda_c(2625)$ and $\Lambda_c(2765)$ resonances} 
\label{sec:charm}

\subsubsection%
          [Predictions of the $\Lambda_c(2595)$ and $\Lambda_c(2625)$ from  the bottom-sector fit]
{\boldmath Predictions of the $\Lambda_c(2595)$ and $\Lambda_c(2625)$ from  the bottom-sector fit}

\begin{figure}[t]
\begin{center}
\makebox[0pt]{\includegraphics[width=0.39\textwidth]{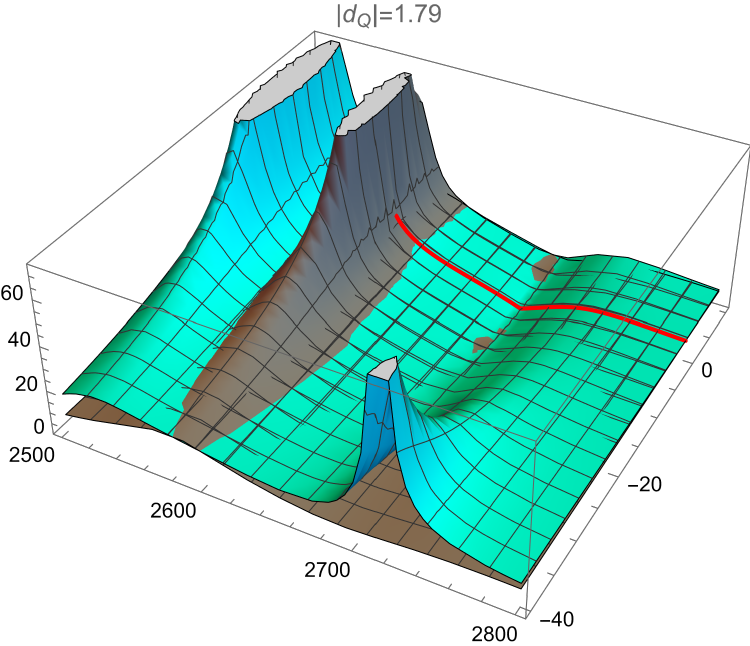}\hspace{0.75cm}\includegraphics[width=0.39\textwidth]{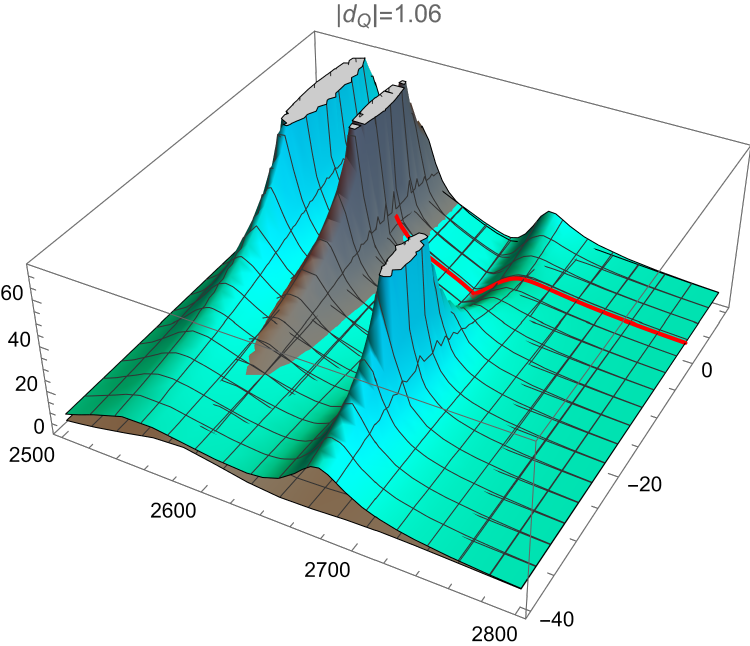}}\\\vspace{0.2cm}
\makebox[0pt]{\includegraphics[width=0.39\textwidth]{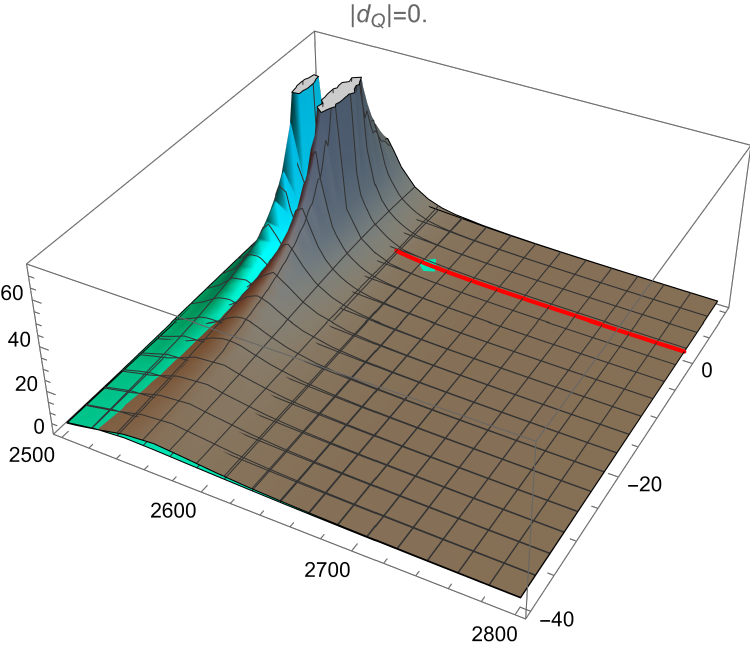}}
\end{center}
\caption{Absolute value of the  $T_{\Sigma_c\pi}$-matrix (in fermi units), both in the first (gray) and second (greenish hues) Riemann sheets and for some of the values of $|d_Q|$ and $\Lambda$ considered in Table~\ref{tab:charm-res}, as a function of complex  $\sqrt{s}=x+iy$ in MeV. We also show the scattering line (red solid line) in all the cases}\label{fig:FRSc}
\end{figure}

From the previous study in the bottom sector, we start by considering the $(|d_Q|=1.79,\, \Lambda=400\,\text{MeV}$) and $(|d_Q|=1.06,\,\Lambda=650\,\text{MeV}$) LEC scenarios. The unitarized amplitudes in the charm sector have also poles in the FRS and SRS which correspond to bound and virtual states and resonances, as can be seen in Fig.~\ref{fig:FRSc} for the $J^P=1/2^-$ spin-parity channel. Results are qualitatively similar for the $T_{\Sigma_c^*\pi}$-matrix $(J^P=3/2^-)$. As it happened in the bottom sector, virtual states have little relevance since they are located behind the bound states and hence their effects are overshadowed by the physical $\Lambda_c(2595)$ and $\Lambda_c(2625)$. However, the SRS pole above the threshold is a QCD state and it affects the $\Sigma_c\pi$ scattering lineshapes (red curve in the plots). As we will see below, this resonance together with its $J^P=3/2^-$ HQSS sibling provides a theoretical explanation for the broad, statistically significant peak seen in the $\Lambda_c\pi^+\pi^-$ spectrum in the region of $2765\,\text{MeV}$~\cite{CLEO:2000mbh}. Note also that the present scheme when the CQM-state exchange potential ($V_{\rm ex}$) is set to zero (bottom panel in Fig.~\ref{fig:FRSc}) produces a virtual state and a bound state, but it does not produce any other near-threshold resonance. 

\begin{table}[t] 
\centering
\resizebox{1.\textwidth}{!}{
\begin{tabular}{cc|cccc|cccc}
\hline\hline
    \multicolumn{2}{c|}{} & \multicolumn{4}{c|}{$\Lambda_c(2595)$}   &\multicolumn{4}{c}{ $\Lambda_c(2625)$} \\  
\hline 
  \multicolumn{2}{c|}{}  & \multicolumn{4}{c|}{$\Sigma_c\pi$ $J^P={\frac{1}{2}}^-$}   & \multicolumn{4}{c}{$\Sigma^*_c\pi$ $J^P={\frac{3}{2}}^-$}     \\
  \hline
 $\Lambda$ [GeV]  & $|d_Q|$  & $M$ [MeV] & $|g_{\Sigma_c\pi}|$ &$P_{\Sigma_c\pi}$ & $\Gamma_{\Lambda_c\pi\pi}^{R}$ [keV] &  $M$ [MeV] & $|g_{\Sigma_c^*\pi}|$ &$P_{\Sigma^*_c\pi}$ & $\Gamma_{\Lambda_c\pi\pi}^{R}$ [keV]  \\
 \hline
$0.4$   &  $1.79\pm 0.11$      &    $2578\pm 3.0$ & $1.72\pm 0.07$  &  $0.49\pm 0.01$ & $60^{+50}_{-30}$ & $2606\pm 2$ & $1.74\pm 0.08$  &$0.27\pm 0.02$  &  $173^{+14}_{-6}$  \\
$0.65$  &  $1.06\pm 0.06$      &    $2581\pm 2.0$ & $1.51\pm 0.04$  &  $0.42\pm 0.02$ & $90^{+60}_{-30}$ & $2608\pm 2$ & $1.32\pm 0.05$  &$0.16\pm 0.01$  &  $118 \pm 8$ \\
$0.9$  &  $0.75\pm 0.04$      &    $2577.7\pm 1.7$ & $1.50\pm 0.02$  &  $0.37\pm 0.01$ & $38^{+19}_{-12}$ & $2608.8\pm 1.7$ & $1.16\pm 0.04$  &$0.12\pm 0.01$  &  $97 \pm 6$   \\
$1.15$    &  $0.55\pm 0.04$      &    $2571.7\pm 1.7$ & $1.51\pm 0.02$  &  $0.31\pm 0.01$ & $8^{+4}_{-3}$ & $2610\pm 2$ & $1.10\pm 0.03$  &$0.11\pm 0.01$  &  $95 \pm 8$    \\
\hline
$1.85\pm 0.04$    &  0      &    $2551.1\pm 1.6$ & $1.35\pm 0.01$  &  $0.18\pm 0.01$ & $0$ & $2615.3\pm 1.6$ & $1.35\pm 0.02$  &$0.18\pm 0.01$  &  $206 \pm 17$ \\
\hline\hline\\[0.5mm]
\end{tabular}
}
\resizebox{1.\textwidth}{!}{
\begin{tabular}{cc|cccc|cccc}
\hline\hline
\multicolumn{2}{c|}{} & \multicolumn{4}{c|}{$\Sigma_c\pi$ [$J^P=1/2^-$]}   & \multicolumn{4}{c}{$\Sigma^*_c\pi$ [$J^P=3/2^-$]}    \\
\hline
 $\Lambda$ [GeV]  & $|d_Q|$  & $M$ [MeV]& $\Gamma$ [MeV] &$|g_{\Sigma_c\pi}|$ & $\phi_{\Sigma_c\pi}$ &  $M$ [MeV] & $ \Gamma$ [MeV] &$|g_{\Sigma^*_c\pi}|$ & $\phi_{\Sigma^*_c\pi}$ \\
 \hline
 $0.4$   &  $1.79\pm 0.11$      &    $2714\pm 6$ & $85.7\pm 0.6$  &  $1.60\pm 0.02$ & $-0.92\pm 0.01$ &  $2754 \pm 6$ & $107.7\pm 0.3$  &$1.80\pm 0.03$ & $-0.77\pm 0.01$  \\
$0.65$  &  $1.06\pm 0.06$      &    $2674\pm 4 $ & $45.2\pm 1.1$ &  $1.33\pm 0.01$& $-0.75\pm 0.01$   & $2711 \pm 3$ & $62.5\pm 0.5$  &$1.66\pm 0.02$ & $-0.57\pm 0.01$ \\
$0.9$  &  $0.75\pm 0.04$      &    $2655\pm 2$ & $20.5\pm 0.9$  &  $0.98\pm 0.02$ & $-0.61\pm 0.01$ &  $2685 \pm 3$ & $31.0\pm 0.8$  &$1.43\pm 0.01$ & $-0.43\pm 0.01$   \\
$1.15$    &  $0.55\pm 0.04$      & $2643\pm 2$ & $8.3\pm 0.8$  &  $0.67\pm 0.03$ & $-0.49\pm 0.01$ &  $2666 \pm 2$ & $12\pm 2 $  &$1.19\pm 0.02$ & $-0.31\pm 0.01$      \\
 \hline \hline
\end{tabular}
}
\caption{Predictions for the $\Lambda_c(2595)$ and $\Lambda_c(2625)$ (top) and  $\Sigma_c^{(*)}\pi$ resonances (bottom) for the different $(\Lambda,|d_Q|)$ pairs considered in  Table~\ref{tab:bottom-res}. In the case of the resonance complex couplings, we give both the absolute value and the phase in radians. For reference, the thresholds used are $(M_{\Sigma_c}+m_\pi)=2593.1\pm 0.5$ MeV and $(M_{\Sigma_c^*}+m_\pi)=2657.6\pm 0.5$ MeV.  }\label{tab:charm-res}
\end{table}

In Table~\ref{tab:charm-res} we compile the properties of the bound states and resonances predicted by the present HQSS scheme in the charm sector and the $(|d_Q|,\Lambda)$ pairs of LECs previously fitted in the bottom sector to the $\Lambda_b(5912)$ and $\Lambda_b(5920)$ states. We see that the present scheme predicts $\Sigma_c^{(*)}\pi$ bound states that can be identified with the $\Lambda_c(2595)$ and $\Lambda_c(2625)$ resonances. The masses are around $15-20\,\text{MeV}$ below the experimental ones, although HQFS and HQSS breaking effects, which can be sizable on the charm mass scale, or ambiguities in the bare mass of the CQM states, could make these differences smaller. We will come back to this below. 

In the CQM calculation of Ref.~\cite{Yoshida:2015tia}, the splitting between the lowest $J^P=1/2^-$ and $3/2^-$ charm states is only of $2\,\text{MeV}$, which is certainly much smaller than the experimental value of around $36\,\text{MeV}$. The disagreement is much worse than in the bottom sector, where the CQM splitting is of $1\,\text{MeV}$ to be compared with the $7\,\text{MeV}$ mass difference between the $\Lambda_b(5912)$ and  $\Lambda_b(5920)$ resonances. Actually, one would expect a larger violation of HQSS in the charm sector in accordance to the trend observed in experiments. The experimental splitting grows by a factor of five from the bottom to the charm sectors, while it only changes by a factor of two in the theoretical CQM of Ref.~\cite{Yoshida:2015tia}. The authors of the latter reference argued that one possible cause for the discrepancy is the coupling to the $ND$ and $ND^*$ baryon-meson scattering states. A simple visual inspection of the spectrum in Fig.~\ref{fig:espectro} might give some support to such an explanation since the $ND$ channel is around $100\,\text{MeV}$ relatively lower in energy than the analog $N\olsi{B}$ one in the bottom sector. Nevertheless, the $ND$ threshold is still more than $200\,\text{MeV}$ above the $\Lambda_c(2595)$. It is more natural to resort to the influence of the closer-in-energy chiral $\Sigma_c\pi$ and $\Sigma_c^*\pi$ channels, which produce a splitting of about $30\,\text{MeV}$, as our results show and can be inferred from the results in Table~\ref{tab:charm-res}. This splitting is in better agreement with the experimentally measured one. Moreover, the interplay between CQM degrees of freedom and the analog $\Sigma_b^{(*)}\pi$ channels in the bottom sector allows us to reproduce the experimental $\Lambda_b(5912)$-$\Lambda_b(5920)$ mass splitting. 

\subsubsection%
          [The $\Lambda_c(2765)$ resonance from   the bottom-sector fit]
{\boldmath The $\Lambda_c(2765)$ resonance from   the bottom-sector fit}
\label{sec:Lc2765}
On the other hand, as mentioned above, in 2001 the CLEO collaboration investigated the spectrum of charmed baryons which decay into  $\Lambda_c\pi^+\pi^-$ and found evidence of a broad state ($\Gamma \approx 50$ MeV) which would have an invariant mass roughly 480 MeV above that of the $\Lambda_c$ ground state baryon~\cite{CLEO:2000mbh}. This is collected in the RPP as the $\Lambda_c(2765)$ or  $\Sigma_c(2765)$ and it is explicitly stated that nothing at all is known about its quantum numbers, including whether it is a $\Lambda_c$ or a $\Sigma_c$, or whether the width might be due to overlapping states~\cite{ParticleDataGroup:2022pth}.  As seen in Table~\ref{tab:charm-res}, for UV cutoffs in the range $400-650\,\text{MeV}$ we obtain broad resonances around $2675-2755\,\text{MeV}$ in both the $J^P=1/2^-$ and $3/2^-$ sectors, which would provide a natural explanation for the excess of events in the $\Lambda_c\pi^+\pi^-$ spectrum reported in Ref.~\cite{CLEO:2000mbh}. These resonances will be heavy quark flavor siblings of those in Table~\ref{tab:bottom-resII}, which we related to the experimental evidence of the $\Lambda_b(6070)$~\cite{LHCb:2020lzx,CMS:2020zzv}. Note that in $J^P=1/2^-$ sector, the $ND$ threshold is around 2800 MeV and it might have a certain impact on the dynamics of a $\Lambda_c$ resonance in the $2700-2750\,\text{MeV}$ region. The right panel of Fig.~\ref{fig:high}  graphically illustrates how would be the resulting spectrum.

\subsubsection%
          [The three-body $\Lambda_c\pi\pi$ decay width and the $g_{\Lambda_c^*\Sigma_c^{(*)}\pi}$ coupling]
{\boldmath The three-body $\Lambda_c\pi\pi$ decay width and the $g_{\Lambda_c^*\Sigma_c^{(*)}\pi}$ coupling}
The $\Gamma_{\Lambda_c\pi\pi}^R$ widths reported in Table~\ref{tab:charm-res} are very small. However, they are unrealistic, in particular those in the $J^P=1/2^-$ channel, because they strongly depend on the resonance masses, when the states are close to the $\Sigma_c^{(*)}\pi $ thresholds, and the poles in Table~\ref{tab:charm-res}  are around 15-20 MeV below the experimental values in both spin-parity sectors. Using the nominal masses of the $\Lambda_c(2595)$ and $\Lambda_c(2625)$, and adding the contribution of neutral and charged pion modes,  the $\Sigma_c^{(*)}-$resonant contribution to  the  $\Lambda_c(2595)$ and  $\Lambda_c(2625)$ decays into $\Lambda_c^+$ and two pions  (see Fig.~\ref{fig:decay}) are~\cite{Nieves:2019nol}
\begin{eqnarray}
\Gamma^{R\dagger}[\Lambda_c(2595)\to \Lambda_c \pi\pi ] &=& (1.9\pm 0.2) \times g^2_{\Lambda_c(2595)\Sigma_c\pi}~{\rm [MeV]}\, ,
 \nonumber \\
 \Gamma^{R\dagger}[\Lambda_c(2625)\to \Lambda_c \pi\pi ] &=& (0.27\pm 0.01) \times g^2_{\Lambda_c(2625)\Sigma_c^*\pi}~{\rm [MeV]}, \label{eq:resulwdth} 
\end{eqnarray}
where the errors account for the variations induced by the uncertainties of the involved hadron masses, for which we take the appropriate experimental masses in each mode: $\Lambda_c(2595)^{+}\to \Sigma_c^{+}(\Lambda_c^+\pi^0)\pi^0$, $\Lambda_c(2595)^{+}\to \Sigma_c^{++}(\Lambda_c^+\pi^+)\pi^-$ and  $\Lambda_c(2595)^{+}\to \Sigma_c^{0}(\Lambda_c^+\pi^-)\pi^+$. For the $\Lambda_c(2625)$ decay, the intermediate $\Sigma_c$ is replaced by the $\Sigma_c^*$ resonance. In the charm sector, these resonant contributions to the $\Lambda_c\pi\pi$ three-body decay channel are much larger than those in the bottom sector because the intermediate $\Sigma_c$ and  $\Sigma_c^*$ states are closer to being on the mass shell, especially for the $\Lambda_c(2595)$ as can be appreciated in Fig.~\ref{fig:espectro}. In this latter case, the hadron mass uncertainties, although small of the order of 0.5 MeV at most, are not negligible compared to the width of the $\Sigma_c$ ($\sim 2 $ MeV), which is on-shell, and produce variations of $\Gamma^{R}[\Lambda_c(2595)]$ of about 10\%. The experimental width of the $\Lambda_c(2595)$ is $2.6 \pm 0.6$ MeV (nearly 100\% saturated by the two $\Lambda_c\pi\pi$ modes), while there exists an upper bound of 0.97 MeV for the $\Lambda_c(2625)$~\cite{ParticleDataGroup:2022pth}. Hence, the experimental $\Gamma[\Lambda_c(2595)]$ provides a direct measurement of the S-wave coupling constant 
\begin{equation}
  g^2_{\Lambda_c(2595)\Sigma_c\pi}=1.37 \pm 0.35,  \label{eq:g2coupl}
\end{equation}
assuming a negligible D-wave contribution~\cite{CDF:2011zbc}. On the other hand, we find that for the coupling constant $h_2$ determined in Ref.~\cite{CDF:2011zbc},
\begin{equation}
    h_2^2=  \frac23\frac{f_\pi^2g^2_{\Lambda_c(2595)\Sigma_c\pi}}{(M_{\Lambda_c(2595)}-M_{\Sigma_c})^2} \times 1.16= 0.35 \pm 0.09\,,
\end{equation}
which agrees with the experimental value of $0.36 \pm 0.08$ fitted to the experimental spectrum in Ref.~\cite{CDF:2011zbc}. The factor 1.16 accounts for the 16\%-difference  between the value of $(1/\Gamma_{\Sigma_c})$, which determines the on-shell three-body decay contribution,  used here (1/1.9 MeV$^{-1}$) to obtain the result in Eq.~\eqref{eq:resulwdth}  and that (1/2.2 MeV$^{-1}$)  employed in Ref.~\cite{CDF:2011zbc}.

The $\pi^0\pi^0$ channel is the 82\%\footnote{This is in excellent agreement with what can be inferred, for the physical $\Lambda_c(2595)$ mass, from Fig. 8 of the experimental analysis of Ref.~\cite{CDF:2011zbc}. This, together with the theoretical understanding of the measured value for $h_2$, supports the hypothesis that in Eq.~(13) of that reference, where there is a typo as pointed out in Ref.~\cite{Nieves:2019nol}, the correct factor $1/\sqrt{2}$ was used in the actual calculations instead of the overall 1/2 that appears there.} and 45\% of the total for the $\Lambda_c(2595)$ and $\Lambda_c(2625)$ partial widths of Eq.~\eqref{eq:resulwdth}, respectively. In the exact isospin limit, the two-neutral-pions partial width is a factor of two smaller than the $\pi^+\pi^-$ one. This isospin breaking is due to the gap of about 6 MeV between the thresholds of $\Sigma_c^{0, ++}\pi^{\pm}$ and $\Sigma_c^+\pi^0$ channels, which becomes important because the  $\Lambda_c(2595)$ would be an extremely shallow bound state. This situation is analog to that of the $X(3872)$, understood as a $J^{PC}=1^{++}$  molecular $D\olsi{D}{}^{\ast}$ state, which will not have a well-defined isospin because of the $D^0\olsi{D}{}^{\ast 0}$-$D^+ D^{\ast -}$ mass difference~\cite{Gamermann:2009fv,Gamermann:2009uq}.  Just as the $X(3872)$ is not merely an isoscalar state and it contains a small mixing with an isovector component~\cite{Hanhart:2011tn,Hidalgo-Duque:2012rqv,Hidalgo-Duque:2013pva,Guo:2013sya}, the $\Lambda_c(2595)$ will also have small isospin one and two components. We have seen that taking into account such components is quite relevant to properly evaluate its three-body $\Lambda_c\pi\pi$ decay width, but they will not qualitatively modify the relation of this resonance with its possible HQSS partner  $\Lambda_c(2625)$ or its bottom siblings $\Lambda_b(5912)$ and $\Lambda_b(5920)$. Such isospin admixtures will have some small quantitative impact in the correct position of the $\Lambda_c(2595)$ pole, and a correct treatment will also require considering isospin breaking terms in the $\Sigma_c\pi$ interaction, and not only in the energy-thresholds. This is beyond the scope of the present study.   

Finally, the upper bound on $\Gamma[\Lambda_c(2625)]$, on the other hand, puts upper limits on the coupling of this resonance to the S-wave $\Sigma_c^*\pi$ pair, but one should bear in mind that in this case, the resonant contribution does not saturate the decay width.

\subsubsection%
          [Molecular content of the $\Lambda_c(2595)$ and $\Lambda_c(2625)$ resonances]%
{\boldmath Molecular content of the $\Lambda_c(2595)$ and $\Lambda_c(2625)$ resonances}
For the $(|d_Q|=1.79,\,\Lambda=400\,\text{MeV}$) and $(|d_Q|=1.06,\,\Lambda=650\,\text{MeV}$) LEC pairs, the $\Sigma_c\pi$-molecular probability of the $\Lambda_c(2595)$ in Table~\ref{tab:charm-res} becomes of the order of $0.4-0.5$, clearly larger than that found for its HQFS sibling $\Lambda_b(5912)$ in the bottom sector. This is because the charmed resonance is located significantly closer to the $\Sigma_Q\pi$ threshold. The molecular content of the $J^P=3/2^-$ $\Lambda_c(2625)$ is smaller and similar to that of the $\Lambda_b(5920)$, confirming that the largest component in the inner structure of this resonance is the compact CQM state. The continuum $\Sigma_c^*\pi$ threshold is about $30\,\text{MeV}$ above and it influences the exact position of this pole and that of the higher broad resonance found around $2750\,\text{MeV}$.

We should note that the molecular probabilities defined in Eq.~\eqref{eq:prob} could depend significantly on the position of the pole through the derivative of the loop function $\left.\partial \olsi{G}_{\Sigma_c^{(*)}\pi}/\partial\sqrt{s}\right|_{\sqrt{s}=\sqrt{s_R}}$. In the $J^P=3/2^-$ sector, this dependence is mild, and the above derivative takes a value of $-0.09$ for the values around $\sqrt{s_R}=2608\,\text{MeV}$ collected in Table~\ref{tab:charm-res} and it only increases to $-0.12$, when $\sqrt{s_R}$ is set to the physical mass of the $\Lambda_c(2625)$. This value for the derivative of the loop function is consistent with the results quoted above for the $\Lambda_b(5912)$ and $\Lambda_b(5920)$ in the bottom sector. 

However the $\Lambda_c(2595)$ is located almost precisely at the $\Sigma_c\pi$ threshold (see Fig.~\ref{fig:espectro}), which changes completely the situation, since the derivative of $ \olsi{G}_{\Sigma_c\pi}$ diverges at the threshold. For the ranges of masses collected in Table~\ref{tab:charm-res}, and using the hadron masses given in Table~\ref{tab:properties}, we find that:\footnote{This value of $-0.18$ is larger than those found in the $J^P=3/2^-$ sector, and it explains the bigger molecular probabilities in the table for $J^P=1/2^-$.}
\begin{subequations}
    \begin{align}
    \left.\partial \olsi{G}_{\Sigma_c\pi}/\partial\sqrt{s}\right|_{\sqrt{s}=2580\,\text{MeV}}   & \simeq -0.18\,,\\
    \left.\partial \olsi{G}_{\Sigma_c\pi}/\partial\sqrt{s}\right|_{\sqrt{s}=M_{\Lambda_c(2595)}}& \simeq -(0.7^{+2.4}_{-0.2})\,.
    \end{align}
\end{subequations} 
To render the molecular probability of Eq.~\eqref{eq:prob} finite, the coupling $g_{\Sigma_c\pi}$ should decrease as the distance (binding) from the state to the threshold is lowered~\cite{Toki:2007ab,Albaladejo:2012te}. Using the result in Eq.~\eqref{eq:g2coupl}, we obtain molecular probabilities for the $\Lambda_c(2595)$ quite close to $1$ and at least larger than $0.6-0.7$. 

In the case of the $\Lambda_c(2625)$, even assuming that the upper bound for the three body $\Lambda_c\pi\pi$ decay width ($\Gamma_{\Lambda_c(2625)}\leqslant 0.97\,\text{MeV}$) were saturated by the $\Sigma_c^*$ resonant mechanism depicted in Fig.~\ref{fig:decay}, the coupling square $g^2 _{\Sigma_c^*\pi}$ would be of the order of 3.5 at most, which might lead to maximum molecular contents for the $\Lambda_c(2625)$ of about 40\%. However, it is experimentally established that $\Lambda_c\pi\pi$ non-resonant mechanisms give rise to the dominant contribution to the total $\Gamma_{\Lambda_c(2625)}$  width~\cite{ARGUS:1993vtm,CLEO:1994oxm}, about $0.5\,\text{MeV}$, which will reduce  $g _{\Sigma_c^*\pi}$ and it will lead to values for $P_{\Sigma_c^*\pi}$ in the range of 10-20\%, in reasonable agreement with the results in Table~\ref{tab:charm-res}.

Therefore, we conclude that the inner structure of the $\Lambda_c(2595)$ and $\Lambda_c(2625)$ would be substantially different, with a dominant $\Sigma_c\pi$ molecular component in the $J^P=1/2^-$ resonance and a CQM compact one in the latter. From this perspective, these two states will not be HQSS partners, as already pointed out in Ref.~\cite{Nieves:2019nol}.

\subsubsection{Heavy quark symmetry breaking effects}

\begin{figure}[t]
\begin{center}
\makebox[0pt]{\includegraphics[width=0.45\textwidth]{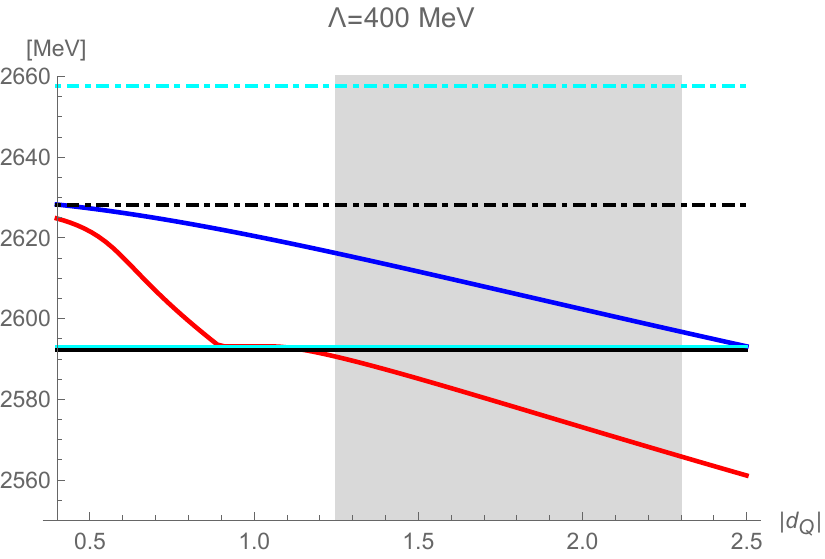}\hspace{0.75cm}\includegraphics[width=0.45\textwidth]{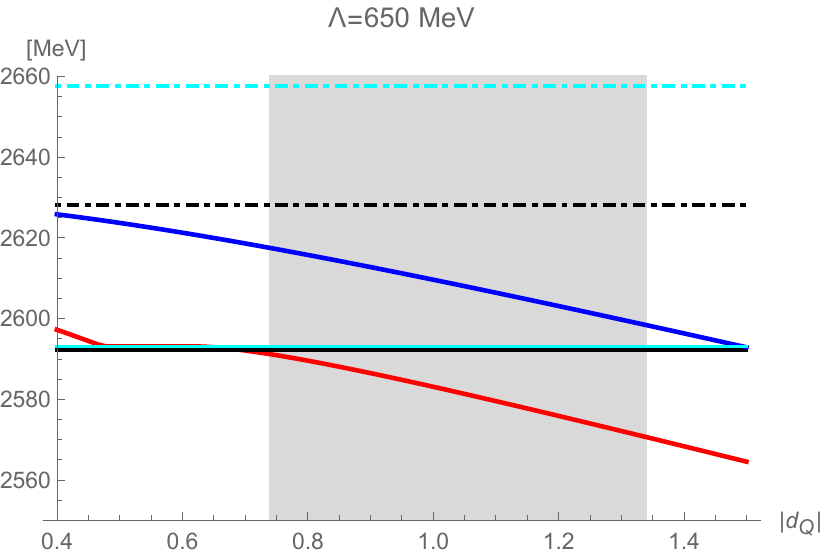}}\\\vspace{1cm}
\makebox[0pt]{\includegraphics[width=0.45\textwidth]{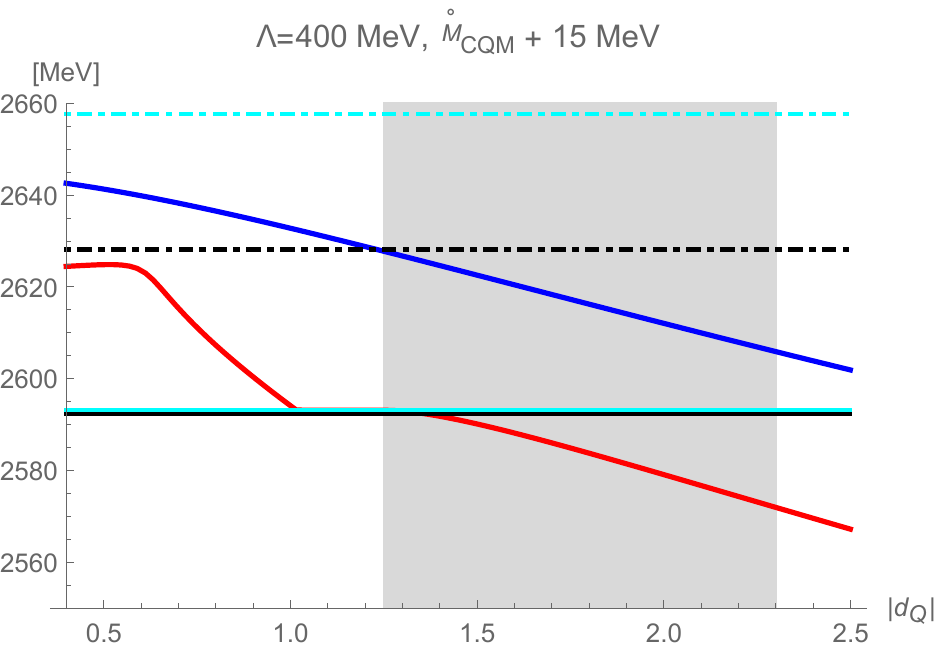}\hspace{0.75cm}\includegraphics[width=0.45\textwidth]{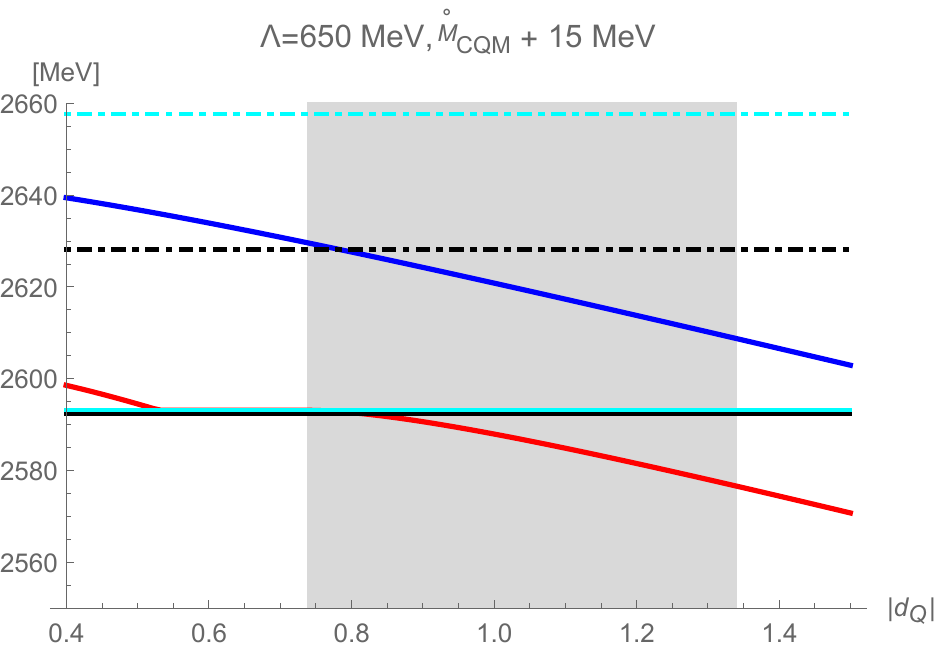}}
\end{center}
\caption{ Top: The solid red (blue) curve shows the real part of the position of the lowest pole of the $T_{\Sigma_c\pi}$ ($T_{\Sigma_c^*\pi}$) matrix as a function of $|d_Q|$ for $\Lambda=400\,\text{MeV}$  (left) and $\Lambda=650\,\text{MeV}$ (right). The pole is always a bound state in the $\frac{3}{2}^-$ sector. For $J^P=\frac{1}{2}^-$,  the pole is located in the FRS or SRS depending if it is below or above the $\Sigma_c\pi$ threshold, except for a region of $|d_Q|$ values, in the interface between both zones,  where the pole is located in the SRS below, but very close to, threshold. The $\Sigma_c\pi$ and  $\Sigma_c^*\pi$ thresholds are displayed by the cyan solid and dashed-dotted lines, respectively. The nominal mass (central value) of the $\Lambda_c(2595)$ ($\Lambda_c(2625)$) resonance is shown by the black (dashed-dotted) solid line. For each UV cutoff, the shaded grey band highlights a $\pm$ 30\% (SU(3) symmetry breaking estimate) variation of $|d_Q|$ with respect to its central value fitted in the bottom sector. Bottom: The same as top panels, but using in the calculations CQM bare masses $15\,\text{MeV}$ greater than those obtained in Ref.~\cite{Yoshida:2015tia}. }\label{fig:charm}
\end{figure}

\begin{figure}[t]\centering
\includegraphics[height=6cm,keepaspectratio]{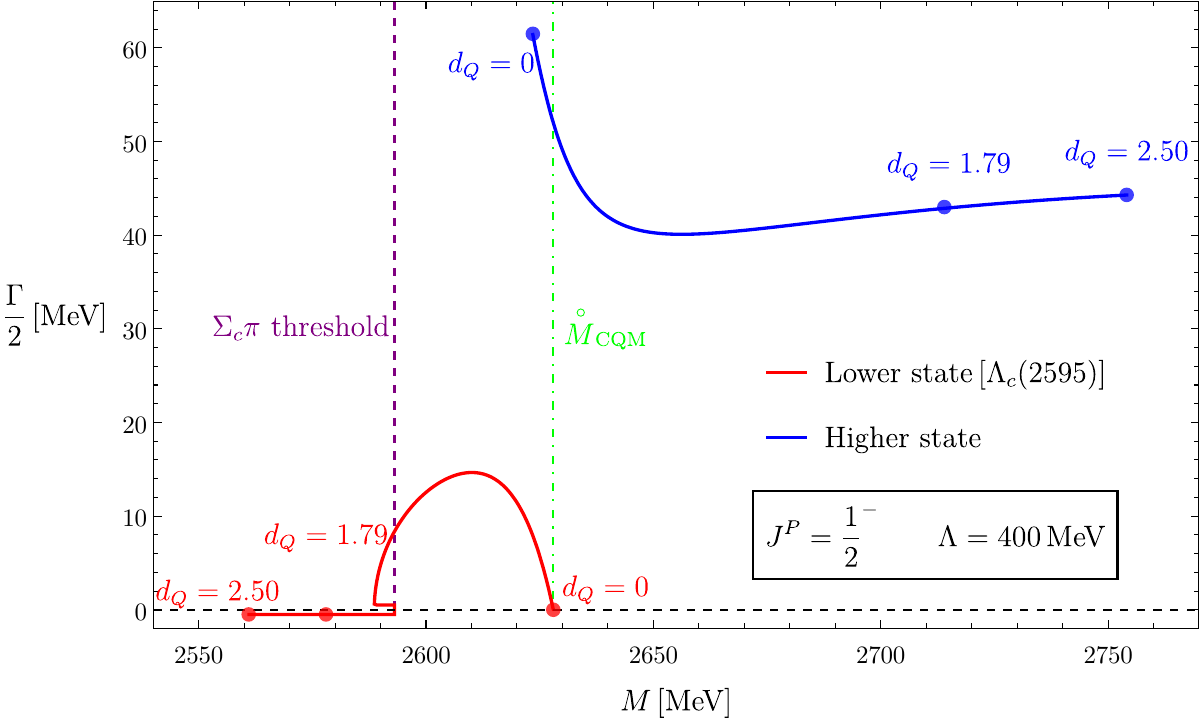}
\caption{Trajectories of the $J^P = \frac{1}{2}^-$ pole positions for $\Lambda = 400\,\text{MeV}$ in the charm sector ($\Sigma_c \pi$ scattering) when the $d_Q$ coupling is varied between $2.5$ and $0$. The blue (red) line represents the higher (lower, associated with $\Lambda_c(2595)$) pole. The higher pole is always a resonance in the SRS, while the lower one transits from a bound state (FRS) to a resonance (SRS) for decreasing values of $d_Q$. The values $d_Q=2.5$, $1.79$, and $d_Q=0$ are highlighted in the trajectories with points. The case $d_Q = 1.79$ is the one used in the first row of the upper and lower parts of Table~\ref{tab:charm-res}. The vertical purple dashed and green dot-dashed lines represent the threshold and the bare CQM mass ($\Mbare = 2628\,\text{MeV}$), respectively. Note that, in the limit $d_Q \to 0$, the lower state goes to $\Mbare$.\label{fig:charm-detail}}
\end{figure}

The above discussions provided a further support for the present theoretical scheme which is based on chiral and heavy quark flavor symmetry, and accounts for the interplay between $\Sigma_Q^{(*)}\pi$ and CQM degrees of freedom. Coming back to the predictions for the $\Lambda_c(2595)$ and $\Lambda_c(2625)$, one should take into account that violations of HQSS and HQFS could be of the order of 20-30\% at the charm quark scale. Such heavy quark symmetry-breaking effects produce variations in the masses of the poles, as can be seen in the top panels of Fig.~\ref{fig:charm} for our preferred $(|d_Q|=1.79,\,\Lambda=400\,\text{MeV}$) and $(|d_Q|=1.06,\,\Lambda=650\,\text{MeV}$) LEC pairs. There, the shaded grey bands show a $\pm$ 30\% (SU(3) symmetry breaking estimate) variation of $|d_Q|$ with respect to its central value fitted in the bottom sector. We see that for the lowest values for $|d_Q|$, the model gives the mass of the $\Lambda_c(2595)$, while the prediction for the $\Lambda_c(2625)$ reaches the region of $2615-2620\,\text{MeV}$, only around $10\,\text{MeV}$ below the experimental value. For $\Lambda=400\,\text{MeV}$ ($\Lambda=650\,\text{MeV}$), if we take $|d_Q|=1.15$ ($|d_Q|=0.69$), we get the mass of the $J^P=1/2^-$ FRS pole at $2592.2\,\text{MeV}$ and  $g_{\Sigma_c\pi}=1$, which leads to $P_{\Sigma_c\pi}\sim 0.7$ in agreement with the discussion above on the dominant molecular nature of the $\Lambda_c(2595)$. In Fig.~\ref{fig:charm-detail}, we further illustrate the dependence  of our results on $d_Q$. There,  we show  the position of the two poles that we find in the $J^P=1/2^-$ sector, for $\Lambda=400\,\text{MeV}$, as the LEC $d_Q (>0)$ decreases from $2.5$ to $0$.\footnote{The pole behaviour analyzed here is very similar to the pion-mass dependence of the $\sigma$ ($f_0(500)$) meson in $\pi\pi$ scattering described in Refs.~\cite{Albaladejo:2012te,Hanhart:2008mx}, to which the reader is referred for further details.} The lower mass pole (red line) for $d_Q=2.5$ is a bound state with a mass around $2560\,\text{MeV}$, and as $d_Q$ decreases it approaches the $\Sigma_c\pi$ threshold, reaching it for $d_Q \sim 1.05$. When $d_Q$ is further reduced, the pole appears in the SRS, at the beginning  in the real axis (virtual pole), but there is a moment ($d_Q \sim 0.97$) for which the pole position gets an imaginary part, with its real part approaching again the threshold. For values of $d_Q$ smaller than around $0.90$, the pole becomes a resonance since its real part is placed above threshold. Now when $d_Q$ is further decreased, the width of this close-to-threshold resonance also decreases,  while its mass approaches that of the bare CQM state, which is obviously reached when $d_Q=0$. The range of values of $d_Q$ between $0.9$ and $1.05$, for which the pole is found in the SRS  below the $\Sigma_c\pi$ threshold, gives rise to the flat behaviour in the left top panel of Fig.~\ref{fig:charm}. The $d_Q-$trajectory of  the higher-pole position is shown by the blue line in Fig.~\ref{fig:charm-detail}. We see that for $d_Q=1.79$ the mass and width of this second state are $2714$ MeV  and  $86$ MeV, respectively. As we discussed in Subsec.~\ref{sec:Lc2765}, this resonance should provide a theoretical explanation to part of the excess of events reported by CLEO in the  $\Lambda_c\pi^+\pi^-$ spectrum  above 2.7 GeV~\cite{CLEO:2000mbh}.

The overall agreement improves when CQM bare masses greater than those given in Ref.~\cite{Yoshida:2015tia} are used. In the lower panels of Fig.~\ref{fig:charm} we show the results obtained by using CQM bare masses increased by $15\,\text{MeV}$, \textit{i.e.} $\Mbare= 2643\,\text{MeV}$ and $2645\,\text{MeV}$ for the $J^P=1/2^-$ and $J^P=3/2^-$ sectors, respectively. It might well be that either one, or both, of the bare CQM masses obtained in Ref.~\cite{Yoshida:2015tia} are not appropriate to describe in the current scheme, where the  $(|d_Q|, \Lambda$) LECs have been fixed to the masses of the $\Lambda_b(5912)$ and $\Lambda_b(5920)$, the dynamics in the charm-sector. This could be because the matching of the CQM of Ref.~\cite{Yoshida:2015tia} and the chiral $\Sigma_Q^{(*)}\pi$ dynamics incorporated in our model might occur at different scales in the bottom and charm sectors. On the other hand, we recall here again the very small mass HQSS splitting predicted in Ref.~\cite{Yoshida:2015tia} for the CQM  $\Lambda_c(2595)$ and $\Lambda_c(2625)$ states. 


\subsection%
          [Strange sector: the $\Lambda(1520)$ and the double pole structure of the $\Lambda(1405)$ resonance]%
{\boldmath Strange sector: the $\Lambda(1520)$ and the double pole structure of the $\Lambda(1405)$ resonance} \label{sec:strange}
In the strange sector, the mixing between the $\lambda-$ and $\rho-$components of the $\Lambda_Q$ CQM  wave function is suppressed by the constituent mass of the strange quark ($\sim$500 MeV), instead of those of the charm or bottom quarks ($\sim 1700$ MeV or $\sim 5100$ MeV, respectively) in the heavy sectors. However, within the CQM of Ref.~\cite{Yoshida:2015tia}, the lowest $1/2^-$ and $3/2^-$ states show still a quite large $\lambda-$dominance in the strange sector.\footnote{For instance in Fig. 10 of that reference, we see that the probability of the $\rho-$mode in the $1/2^-$ lowest $\Lambda$ state is well below 5\%.}  For this reason, we assigned the $\lambda-$excitation mode (total angular momentum and parity of the light degrees of freedom coupled to $j_q^\pi=1^-$) to the first CQM $1/2^-$ and $3/2^-$ states  in Fig.~\ref{fig:espectro}. Furthermore, we expect that the scheme presented in the previous sections for the bottom and charm sectors, where the $\Sigma^*_Q\pi$ chiral pair with $j_q^\pi=1^-$ is coupled to the lowest $\lambda-$CQM state, should work reasonably well to describe the $J^P=3/2^-$ $\Lambda(1520)$ state. Nevertheless, we expect sizable HQSS breaking corrections in the numerical values of the $|d_Q(\Lambda)|$ LEC, which might differ from those found in the heavy sectors.  To describe the dynamics of the next resonance in this sector, \textit{i.e.} $\Lambda(1690)$, it would be necessary to include the second CQM level, which is located on top of the physical resonance and is predominantly a $\rho-$mode excitation.\footnote{Note in Fig.~\ref{fig:espectro} that the first two CQM $3/2^-$ states are separated by 150 MeV, 290 MeV and 330 MeV in the strange, charm and bottom sectors, respectively~\cite{Yoshida:2015tia}. Hence, we see that this $\rho-\lambda$ mode splitting remains still quite sizable at the strange quark mass. On the other hand, the energy separation between the second and the third states increases notably when the mass of the spectator quark $Q$ decreases and it goes from  12 MeV for bottom states to 39 MeV for charm states and finally to 125 MeV for the strange ones~\cite{Yoshida:2015tia}. This would be the splitting between the $(\ell_\rho=1)$ excitations with {\it ldof} coupled to $j_q^\pi=1^-$ and $2^-$ in the heavy sectors. Clearly, HQSS is considerably broken at the strange scale for these two states, and therefore these two CQM levels should be mixture of the mathematical ones with well-defined $j_q^\pi$ quantum numbers.} Therefore, the study of the second resonance $\Lambda(1690)$ in the strange sector is beyond the scope of this work. This contrasts to what we found in the bottom and charm sectors for which we were able to discuss also the $\Lambda_b(6070)$ and the $\Lambda_c(2765)$, since they were located at least $150\,\text{MeV}$ below the CQM $\rho-$type levels. 

In what respects the $J^P=1/2^-$ sector, as we discussed through Secs.~\ref{sec:intro} and \ref{sec:tmatrix} and seen in Fig.~\ref{fig:espectro}, it is necessary to include the chiral coupled-channel dynamics of the $\Sigma\pi$ and $N\olsi{K}$ pairs. The $\lambda-$CQM level should play a minor role in the double pole associated to the $\Lambda(1405)$, with the LEC $c_s$, introduced in Eq.~\eqref{eq:leccs}, accounting for the coupling of this CQM state and the $N\olsi{K}$ meson pair. However, the $\rho-$CQM pole would be relevant for the second physical state $\Lambda(1670)$.

\subsubsection%
          [The $\Lambda(1520)$ state]
{\boldmath The $\Lambda(1520)$ state}

\begin{table}[t] 
\centering
\begin{tabular}{cc|ccc}
\hline\hline
\multicolumn{2}{c|}{} & \multicolumn{3}{c}{$\Sigma^*\pi$ [$J^P=3/2^-$]}    \\
\hline
 $\Lambda$ [GeV]  & $|d_s|$  & $M$ [MeV]& $P_{\Sigma^*\pi}$  &$|g_{\Sigma^*\pi}|$ \\
 \hline
 $0.4$   &  $1.79\pm 0.11$      &     $1498.3 \pm 2.8$ & $0.35\pm 0.1$  &$1.83\pm 0.07$  \\
$0.65$  &  $1.06\pm 0.06$      &    $1502.9_{-1.4}^{+2.3}$ & $0.26\pm 0.1$  &$1.51\pm 0.04$  \\
$0.9$  &  $0.75\pm 0.04$      &     $1503.1 \pm 1.9 $ & $0.24\pm 0.1$  &$1.44\pm 0.02$    \\
$1.15$    &  $0.55\pm 0.04$      &   $1501.8 \pm 2.0$ & $0.23\pm 0.01 $  &$1.44\pm 0.01$      \\
 \hline
$1.85\pm 0.04$    &  0      &$1493.5 \pm 1.4$ & $0.21\pm 0.01 $  &$1.48\pm 0.02$   \\
 \hline\hline
 \end{tabular}
\caption{Properties of the $\Sigma^*\pi$ bound state (FRS pole below threshold)  predicted in the strange sector for the different UV cutoffs contemplated in  Table~\ref{tab:bottom-res}. For reference, the threshold used in this work is $(M_{\Sigma*}+m_\pi)=1524.6\pm 2$ MeV, while the mass and width of the $(J^P=3/2^-)$ $\Lambda(1520)$ resonance compiled in the RPP~\cite{ParticleDataGroup:2022pth} are $(1519\pm 1)$ MeV and $(16\pm 1)$ MeV, respectively. The uncertainty of 2 MeV in the threshold energy has not be taken into account in the masses quoted in the table.\label{tab:strange-res32}}
\end{table}

\begin{figure}[t]
\begin{center}
\makebox[0pt]{\includegraphics[width=0.3\textwidth]{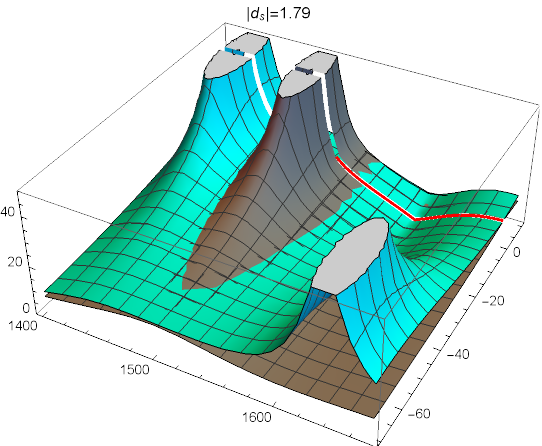}\hspace{0.75cm}\includegraphics[width=0.3\textwidth]{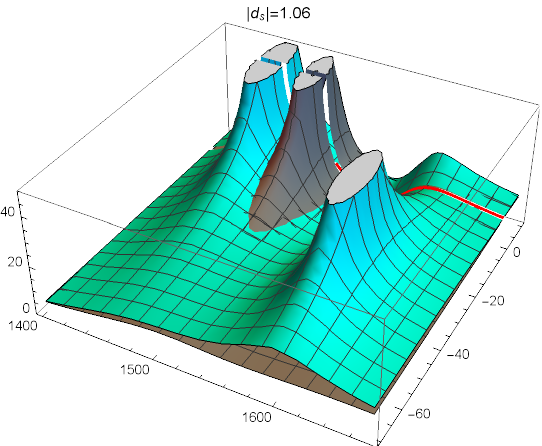}}\\\vspace{0.4cm}
\makebox[0pt]{\includegraphics[width=0.3\textwidth]{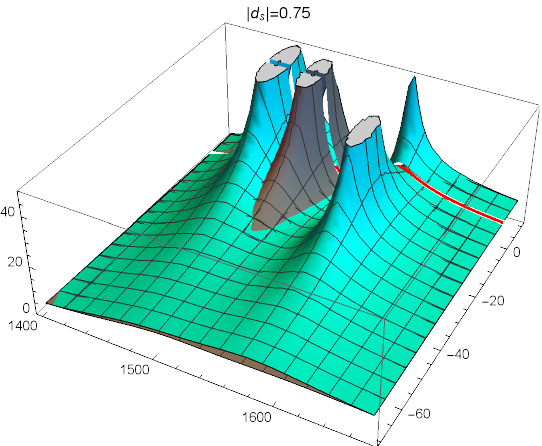}\hspace{0.75cm}\includegraphics[width=0.3\textwidth]{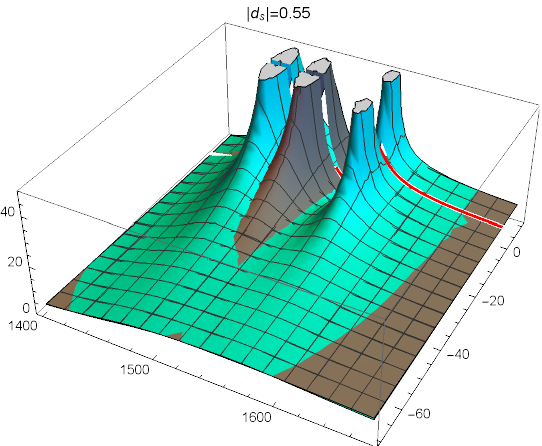}\hspace{0.75cm}\includegraphics[width=0.3\textwidth]{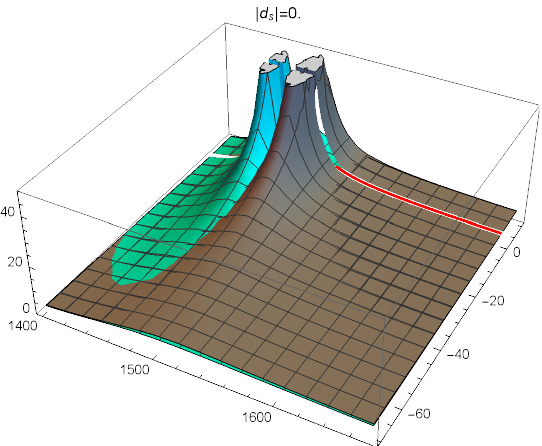}}
\end{center}
\caption{Absolute value of the  $T_{\Sigma^*\pi}$-matrix (in fermi units), both in the first (gray) and second (greenish hues) Riemann sheets and for the values of $|d_s|$ and $\Lambda$ considered in Table~\ref{tab:strange-res32}, as a function of  complex  $\sqrt{s}=x+iy$ in MeV. We also show the scattering line (red solid line) in all the cases.\label{fig:FRSs32}}
\end{figure}

\begin{figure}[t]
\begin{center}
\makebox[0pt]{\includegraphics[width=0.4\textwidth]{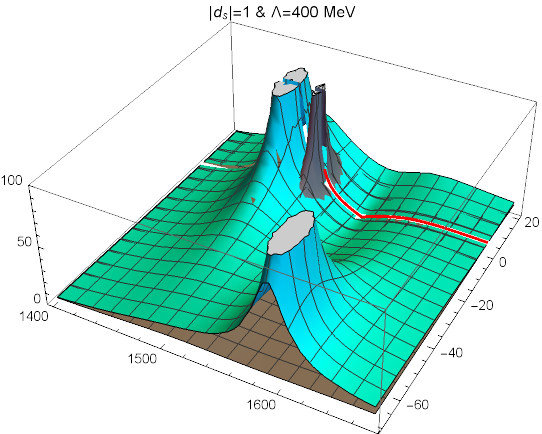}\hspace{1.cm}\includegraphics[width=0.4\textwidth]{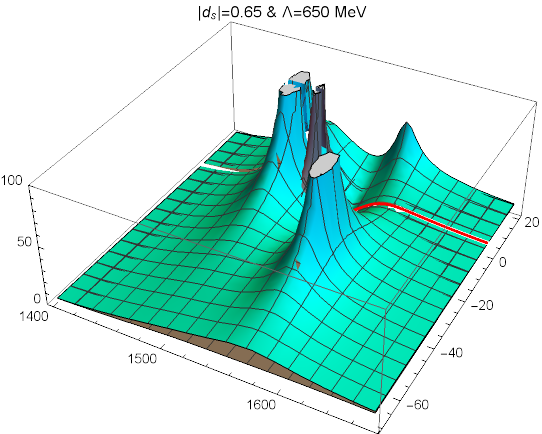}}
\end{center}
\caption{Absolute value of the  $T_{\Sigma^*\pi}$-matrix (in fermi units), both in the first (gray) and second (greenish hues) Riemann sheets and for  $(|d_Q|=1.0,\,\Lambda=400\,\text{MeV}$ (left) and $(|d_Q|=0.65,\,\Lambda=650\,\text{MeV}$ (right), as a function of  complex  $\sqrt{s}=x+iy$ in MeV.  The FRS pole is placed at 1518 MeV in both cases, while the above threshold SRS ones are located at  $(M,\Gamma)=(1590, 115)\,\text{MeV}$ and $(M,\Gamma)=(1571, 60)\,\text{MeV}$, respectively.\label{fig:FRSs32-bis}}
\end{figure}

From the previous study in the bottom sector, we consider five different scenarios characterized by pairs of LECs: $(|d_s|=1.79,\,\Lambda=400\,\text{MeV})$, $(|d_s|=1.06,\,\Lambda=650\,\text{MeV})$, $(|d_s|=0.75,\,\Lambda=900\,\text{MeV})$, $(|d_s|=0.55,\,\Lambda=1150\,\text{MeV})$ and $(|d_s|=0,\,\Lambda=1850\,\text{MeV})$. The $\Sigma^*\pi$ unitarized amplitude has also poles in the FRS and SRS, as can be seen in Fig.~\ref{fig:FRSs32}. From the results collected in Table~\ref{tab:strange-res32}, we see that the pole in the FRS is found in the range $1494-1507\,\text{MeV}$,\footnote{Here we also add the uncertainty of $2\,\text{MeV}$ in the mass of the $\Sigma^*$ which is not considered in the table.} when the interplay between chiral and CQM degrees of freedom are taken into account, and at $1493.5\,\text{MeV}$ when $|d_s|$ is set to zero. In addition, the three-body decay $\Gamma_{\Lambda\pi\pi}^R$ width, discussed in Subsec.~\ref{sec:observables} takes small values in the range $2.0-2.5\,\text{MeV}$. This state should be identified with the $\Lambda(1520)$ resonance, which has a nominal mass of $1519 \pm 1\,\text{MeV}$ and a width of $16\pm 1\,\text{MeV}$. Around 90\% of its width is due to the D-wave open channels $N\olsi{K}$ and $\Sigma\pi$, not considered in the present approach.  


Coming back to Fig.~\ref{fig:FRSs32}, we also see that for higher values of $\Lambda$ there appears a second resonance, which approaches the energy region of $1550\,\text{MeV}$ and becomes narrower and narrower as the UV cutoff increases.\footnote{For this second pole, we find $(M,\Gamma)=$ $(1634, 113)$, $(1595, 65)$, $(1574, 33)$, and $(1559, 15)\,\text{MeV}$ for $\Lambda=400$, $650$, $900$, and $1150\,\text{MeV}$, respectively.} This state would be visible when it is narrow enough, however it has no counterpart in the RPP~\cite{ParticleDataGroup:2022pth}. Based on this discussion, we conclude that it is reasonable to assume that UV cutoffs in the $400-650\,\text{MeV}$ range provide the most sensible results.

For $\Lambda= 400\,\text{MeV}$ ($\Lambda=650\,\text{MeV}$), we can tune $|d_s|= 1.0$ ($\,|d_s|=0.65$) to place the $\Sigma^*\pi$ bound state, which we identify with the $\Lambda(1520)$, at $1518\,\text{MeV}$, with $|g_{\Sigma^*\pi}|=1.25\, (1.19)$, a molecular probability $P_{\Sigma^*\pi}\sim 0.3$ and a three-body decay width $\Gamma_{\Lambda\pi\pi}^R\sim 4.1~ (3.7)\,\text{MeV}$. For these LECs, a second broad  state (see Fig.~\ref{fig:FRSs32-bis}) also appears  at  $(M,\Gamma)=(1590, 115)\,\text{MeV}$ and $(M,\Gamma)=(1571, 60)\,\text{MeV}$ for $\Lambda=400\,\text{MeV}$  and $\Lambda=650\,\text{MeV}$, respectively. The effects of this additional resonance might be difficult to observe and yet it could have some influence in the dynamics of the higher $\Lambda(1690)$ $3/2^-$ state, which should also have a sizable D-wave $N\olsi{K}$ and $\Sigma\pi$ components~\cite{ParticleDataGroup:2022pth} and a significant inner CQM structure driven by a $\rho-$excitation mode (see Fig.~\ref{fig:espectro}).

\subsubsection%
          [Pole structure of the $\Lambda(1405)$ region]
{\boldmath Pole structure of the $\Lambda(1405)$ region}
\begin{table}[t] 
\centering
\begin{tabular}{cc|cccc|cccc}
\hline\hline
\multicolumn{2}{c|}{} & \multicolumn{4}{c|}{pole 1  }   & \multicolumn{4}{c}{pole 2  }    \\
\hline
 $d^\prime_s$ & $c_s$  & $M$ [MeV] & $\Gamma$ [MeV] & $g_{\Sigma\pi}$ &  $g_{N\olsi{K}}$  &  $M$ [MeV] & $ \Gamma$ [MeV] &$g_{\Sigma\pi}$ &  $g_{N\olsi{K}}$\\
 \hline
 0 & 0 & 1437.7$^*$& 2.5$^*$& $(0.2 - 0.7i)$ &$(1.2 + 1.1i)$ & 1369.0 & 142.4 & $(2.1 - 1.5 i)$ & $(1.0 - 1.0 i) $\\ 
 \hline
 1.0 & $-1$ & 1441.0$^*$ & 4.7$^*$ & $(0.4 - 0.9i)$ & $(1.4 + 1.3i)$ & 1356.3 & 111.6 &$(2.2 - 1.4 i)$ & $(1.3 - 1.0 i) $   \\
 1.0 & 0      & 1436.7$^*$ & 0.0$^*$ & $(0.4 + 0.4i)$ & $(1.1 - 1.0i)$ & 1354.9 & 116.5 &$(2.2 - 1.4 i)$ & $(1.0 - 0.8 i) $  \\
  1.0 & 1     & 1433.2\phantom{$^*$} &2.3\phantom{$^*$} & $(0.0 - 0.3i)$ & $(1.4 + 0.3i)$ & 1353.8 & 120.2 &$(2.1 - 1.5 i)$ & $(0.8 - 0.6 i) $  \\
  1.0 & 2     & 1425.0\phantom{$^*$} &0.9\phantom{$^*$} & $(0.0 +0.2i)$ & $(2.1 + 0.0i)$ & 1353.0 & 122.8 &$(2.1 - 1.5 i)$ & $(0.5 - 0.4 i) $  \\ 
\hline\hline 
\end{tabular}
\caption{ Properties of the poles found in the strange $J^P=(1/2)^-$ sector in the  $\Lambda(1405)$ region for $\Lambda= 400$ MeV and different values of the LECs $d^\prime_s$ and $c_s$. $^*$: The pole 1 is virtual since it is above $N\olsi{K}$ threshold ($1434.52$ MeV), but located in the (10) Riemann sheet instead of the (11). The two poles in the RPP are $\left(1421^{+3}_{-2}{}^{+16}_{-20},38^{+16}_{-10}\pm 20\right)$ and $\left(1388\pm 9{}^{+11}_{-78}, 230\pm 50{}^{+0}_{-140}\right)$~\cite{ParticleDataGroup:2022pth}. } \label{tab:strange12}
\end{table}

\begin{figure}[t]
\begin{center}
\makebox[0pt]{\includegraphics[width=0.4\textwidth]{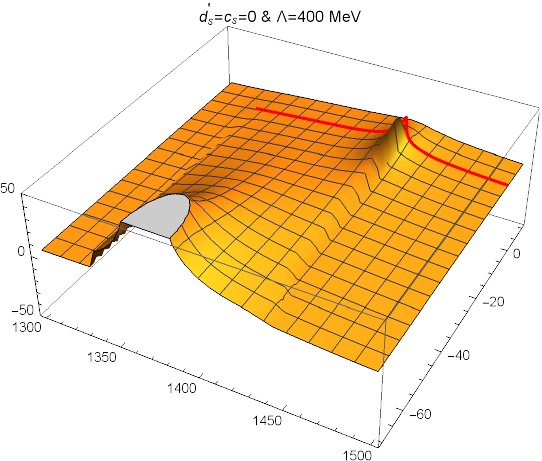}\hspace{1.cm}\includegraphics[width=0.4\textwidth]{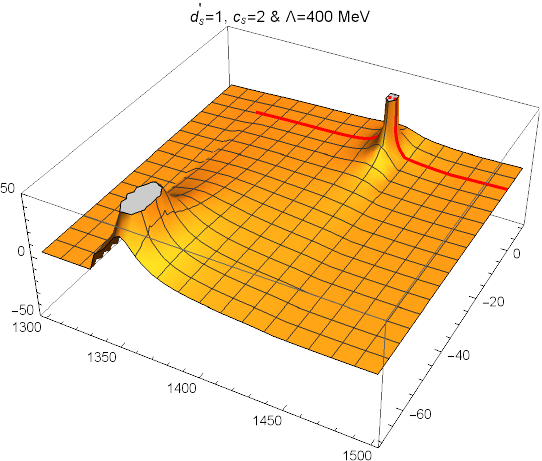}}
\end{center}
\caption{Absolute value of the $[J^P=(1/2)^-]$ $T_{N\olsi{K}\to\Sigma\pi}$ matrix element (in fermi units) for both the FRS [Im$\sqrt{s}>0$] and the SRS [Im$\sqrt{s}<0$] as a function of the complex  $\sqrt{s}=x+iy$ in MeV. The UV cutoff is $\Lambda=400$ MeV, and CQM degrees of freedom are disconnected in the left plot, while they are coupled to the baryon-meson pairs in the right panel using  $d_s^\prime=1$ and $c_s=2$. We also display the scattering line (red solid curve) in both cases. As noted in  Table~\ref{tab:strange12}, in the left panel, the higher $\Lambda(1405)$ pole is related to a virtual state which produces peaks in the FRS and SRS. In the right panel, the higher $\Lambda(1405)$ shows up as a narrow resonance (pole in the SRS) close, but below, the $N\olsi{K}$ threshold. }\label{fig:FRSs12}
\end{figure}

From the discussion in the previous subsection, we have seen that $\Lambda= 400$ MeV and $|d_s|= 1.0$ provide a reasonable description of the  $J^P=3/2^-$ $\Lambda(1520)$ resonance. In the  $J^P=\frac{1}{2}^-$ sector, in principle we should consider the  $\Sigma\pi-N\olsi{K}$  chiral coupled-channels and the couplings $d^\prime_s$ and $c_s$ of these two hadron/meson-baryon pairs to the CQM bare state (Eq.~\eqref{eq:leccs}). We fix the UV cutoff to $400\,\text{MeV}$ and examine the obtained spectrum for $d^\prime_s=|d_s|= 1.0$ and different values of the LEC $c_s$ in the interval $[-1,2]$. Results are compiled in Table~\ref{tab:strange12} and illustrated for two sets of the LECs in Fig.~\ref{fig:FRSs12}.

As expected from the discussion of Fig.~\ref{fig:espectro}, the main features of the two-pole structure in the  $\Lambda(1405)$ region are provided by the chiral $\Sigma\pi-N\olsi{K}$ coupled-channel dynamics (see the related note in the RPP~\cite{ParticleDataGroup:2022pth}),\footnote{The importance of chiral dynamics can be seen in other two-pole/two-state structures, such as the $D^{\ast}_0(2400)$ \cite{Albaladejo:2016lbb,Du:2017zvv} or the $K_1(1270)$ \cite{Roca:2005nm,Geng:2006yb,Garcia-Recio:2010enl}. See also Ref.~\cite{Xie:2023cej}.} which is barely affected by the interplay with the CQM degrees of freedom. Indeed, a qualitatively correct picture is found without taking into account the CQM state, whose largest effect is driven by its coupling $c_s$ to the $N\olsi{K}$ pair, since the $\Sigma\pi$ threshold is around $200\,\text{MeV}$ below $\Mbare$ from Ref.~\cite{Yoshida:2015tia}.  

In the right panel of Fig.~\ref{fig:FRSs12}, we observe a typical two-state structure, obtained considering the interplay between chiral and CQM degrees of freedom, for which the numerical values of resonance masses, widths and couplings have been compiled in the last row of Table~\ref{tab:strange12}.  There appears a broad resonance, located well below the $N\olsi{K}$ threshold and with a dominant coupling to the $\Sigma \pi$ open channel. The second state close to, but below, the $N\olsi{K}$ threshold is quite narrow because it has a significantly smaller coupling to $\Sigma \pi$. When the CQM effects are switched off (left panel of the figure and first row of the table), the mass and width of the wide state qualitatively hardly change, while the narrow state is a bit more affected and becomes virtual, being its pole found in an adjacent Riemann sheet. Nevertheless, we can clearly see the effect of this pole in the scattering lineshape which is shown by a red curve in the 3D plot of the left panel of  Fig.~\ref{fig:FRSs12}.   Most of the studies on the $\Lambda(1405)$ region within the UChPT framework consider the higher-energy $\Lambda\eta$ and $\Xi K$ channels whose sub-threshold contributions widen the state and play a non-negligible role in placing its pole in the proper Riemann sheet (see  Subsec.~\ref{sec:uchpt}).

%

	\section{Conclusions}\label{sec:concl}
We have presented a detailed study of the lowest-lying ${\frac{1}{2}}^-$ and ${\frac{3}{2}}^-$ $\Lambda_{Q}$ resonances both in the heavy (bottom and charm) and the strange quark sectors. We have paid special attention to the interplay between CQM and chiral baryon-meson degrees of freedom, which are coupled using a unitarized scheme consistent with LO heavy quark symmetries. The main conclusions of this work, which could be inferred from the spectrum depicted in Fig.~\ref{fig:espectro}, are: 
\begin{enumerate}
\item The $\Lambda(1405)$ chiral two-pole pattern does not have analog in the $\frac12^-$ charmed and bottom sectors, because the $ND^{(*)}$ and $N\olsi{B}{}^{(*)} $ channels do not play for heavy quarks the decisive role that the $N \olsi{K}$ does in the strange sector, and also due to the notable influence of the bare CQM states for the charm and bottom resonances. 

\item  The $\Lambda_b(5912)$, $\Lambda_b(5920)$ and the $\Lambda_c(2625)$, and the $\Lambda(1520)$ admitting  larger breaking corrections,  are heavy-quark spin-flavor siblings. They can be seen as dressed CQM states with $\Sigma_{(Q)}^{(*)}\pi$ molecular components and couplings of the order of 30\% and 1.5, respectively. Nevertheless, we have shown the importance of taking into account the chiral $ \Sigma_{c,b}^{(*)}\pi$  channels to obtain a good description of these resonances.

\item The ${J^P=\frac{1}{2}}^-$ $\Lambda_c(2595)$ has, however, a higher molecular probability of at least $50\%$, and even values greater than $70\%$ can be easily accommodated. This is because it is located almost on top of the threshold of the $\Sigma_c\pi$ chiral pair, which largely influences its properties. The mass of the bare three-quark state is $30\,\text{MeV}$ above. Although the {\it ldof} in this resonance would be coupled to $j_q^\pi=1^-$ as in the $\Lambda_b(5912)$, $\Lambda_b(5920)$, and $\Lambda_c(2625)$ cases, the $\Lambda_c(2595)$ should not be considered a heavy-quark spin-flavor partner of the former ones.

The isospin breaking due to the gap of about 6 MeV between the thresholds of $\Sigma_c^{0,´++}\pi^{\pm}$ and $\Sigma_c^+\pi^0$ channels becomes important to describe the width of the $\Lambda_c(2595)$, which turns out to be an extremely shallow bound state. This situation is analog to that of the $X(3872)$, understood as a $J^{PC}=1^{++}$  molecular $D\olsi{D}{}^*$ state, which will not have a well defined isospin because of the $D^0\olsi{D}{}^{*0}-D^+D^{*-}$ mass difference~\cite{Gamermann:2009uq}. Just as the $X(3872)$ is not merely an isoscalar state and it contains a small mixing with an isovector component~\cite{Hanhart:2011tn,Hidalgo-Duque:2012rqv,Hidalgo-Duque:2013pva,Guo:2013sya}, the $\Lambda_c(2595)$ will also have small isospin one and two components. We have seen that taking into account such components is quite relevant to properly evaluate its three-body $\Lambda_c\pi\pi$ decay width, but they will not qualitatively modify the relation of this resonance with its possible HQSS partner  $\Lambda_c(2625)$ or bottom siblings $\Lambda_b(5912)$ and $\Lambda_b(5920)$. 

\item We have predicted bottomed and charmed higher-energy $J^P=1/2^-$ and  $J^P=3/2^-$ resonances. These states would contribute to account for the broad excesses of events observed by the LHCb~\cite{LHCb:2020lzx} and CMS~\cite{CMS:2020zzv} in the region of $6040-6100\,\text{MeV}$ and by CLEO~\cite{CLEO:2000mbh} around 2765 MeV in the $\Lambda_b^0\pi^+\pi^-$ and $\Lambda_c^+\pi^+\pi^-$ spectra, respectively. In this way, we have predicted the existence of two  $\Lambda_b(6070)$ and two $\Lambda_c(2765)$ HQSS \& HQSF sibling odd parity states.

\end{enumerate}

 Finally, we would like to stress that ignoring the interplay between hadron and bare quark-model degrees of freedom, which is quite common in both hadron-molecular and CQM approaches, cannot be in most of the cases safely done. Here, we have illustrated how the inner structure and finer details   of the lowest-lying ${\frac{1}{2}}^-$ and ${\frac{3}{2}}^-$ $\Lambda_{Q}$  turn out to be a combined result of both types of physical pictures, with the renormalization procedure becoming a fundamental ingredient because of the non-perturbative character of the hadron spectrum.   

	\section*{Acknowledgements}
This research has been supported  by the Spanish Ministerio de Ciencia e Innovaci\'on (MICINN) and the European Regional Development Fund (ERDF) under contracts PID2020-112777GB-I00, the EU STRONG-2020 project under the program H2020-INFRAIA-2018-1, grant agreement no. 824093 and by  Generalitat Valenciana under contract PROMETEO/2020/023. M.~A. acknowledges finanacial support from Generalitat Valencia (GVA) through Grant no. CIDEGENT/2020/002. M. A. and A. F. thank the warm support of ACVJLI.

	

	


	
	

\end{document}